\documentclass[trackchanges,twocolumn]{aastex701}
\usepackage{amsmath}
\usepackage{placeins}
\usepackage{graphicx}   
\usepackage{subcaption}
\usepackage{hyperref}

\begin{document}

\title{RV and TTV Measurements of Two Transiting Long-Period Giants around TOI-4600}

\author[orcid=0009-0008-0606-9873,sname='Hu']{Tong Hu}
\affiliation{Department of Astronomy, Tsinghua University, Beijing 100084, China}
\email{hutong0516@google.com}  

\author[orcid=0000-0001-5695-8734,sname='Lin']{Zitao Lin}
\affiliation{Department of Astronomy, Tsinghua University, Beijing 100084, China}
\email{lztssealevel@gmail.com}

\author[orcid=0000-0002-6937-9034,sname='Wang']{Sharon X. Wang}
\affiliation{Department of Astronomy, Tsinghua University, Beijing 100084, China}
\email{sharonw@tsinghua.edu.cn}

\author[0000-0003-3015-6455]{Mu-Tian Wang}
\affiliation{School of Astronomy and Space Science, Nanjing University, Nanjing 210023, China.}
\affiliation{Key Laboratory of Modern Astronomy and Astrophysics, Ministry of Education, Nanjing, 210023, People’s Republic of China}
\email{mutianwang@smail.nju.edu.cn}

\author[orcid=0000-0002-4510-2268]{Ismael Mireles}
\affiliation{Department of Physics and Astronomy, The University of New Mexico, 210 Yale Blvd NE, Albuquerque, NM 87106, USA}
\email{mirelesi@unm.edu}

\author[orcid=0000-0003-4733-6532]{Jacob Bean}
\affiliation{Department of Astronomy \& Astrophysics, University of Chicago, Chicago, IL 60637, USA}
\email{jacobbean@uchicago.edu}

\author[orcid=0000-0003-2404-2427]{Madison Brady}
\affiliation{Department of Astronomy \& Astrophysics, University of Chicago, Chicago, IL 60637, USA}
\email{madisontbrady@gmail.com}

\author[orcid=0009-0003-1142-292X]{Nina Brown}
\affiliation{Steward Observatory, University of Arizona, Tucson, AZ 85721, USA}
\email{ninambrown@arizona.edu}

\author{Qikang Feng}
\affiliation{Department of Astronomy, School of Physics, Peking University, Beijing 100871, People’s Republic of China
Kavli Institute for Astronomy and Astrophysics, Peking University, Beijing 100871, People’s Republic of China}
\email{2201110285@stu.pku.edu.cn}

\author[orcid=0000-0002-4503-9705]{Tianjun Gan}
\affiliation{Department of Astronomy, Westlake University
Hangzhou, 310030, China}
\email{tianjungan@gmail.com}

\author[orcid=0009-0005-3557-5183]{Chengyang Ji}
\affiliation{Department of Astronomy, Tsinghua University, Beijing 100084, China}
\email{jicy23@mails.tsinghua.edu.cn}

\author[orcid=0009-0007-4501-4376]{Xue Li}
\affiliation{School of Astronomy and Space Science, University of Chinese Academy of Sciences, Beijing 100049, People's Republic of China}
\email{lixue@bao.ac.cn}

\author[orcid=0009-0002-5981-8520]{Jiayue Zhang}
\affiliation{Department of Astronomy, Tsinghua University, Beijing 100084, China}
\email{zhangjy23@mails.tsinghua.edu.cn}

\author[orcid=0000-0003-4508-2436]{Ritvik Basant}
\affiliation{Department of Astronomy \& Astrophysics, University of Chicago, Chicago, IL 60637, USA}
\email{rbasant@uchicago.edu}

\author[orcid=0000-0002-4070-7831]{Nikita Chazov}
\affiliation{Kourovka observatory \textbar \ Ural Federal University, 19 Mira street, Yekaterinburg, 620002, Russia}
\email{nikita.chazov@urfu.ru}

\author[orcid=0000-0002-9003-484X]{David Charbonneau}
\affiliation{Center for Astrophysics \textbar\ Harvard \& Smithsonian, 60 Garden Street, Cambridge, MA 02138, USA}
\email{dcharbonneau@cfa.harvard.edu}

\author[orcid=0000-0001-6588-9574]{Karen~A.~Collins} 
\affiliation{Center for Astrophysics ${\rm \mid}$ Harvard {\rm \&} Smithsonian, 60 Garden Street, Cambridge, MA 02138, USA}
\email{karen.collins@cfa.harvard.edu}

\author[orcid=0009-0005-1486-8374]{Tanya Das}
\affiliation{Department of Astronomy \& Astrophysics, University of Chicago, Chicago, IL 60637, USA}
\email{tanyadas@uchicago.edu}

\author[orcid=0000-0003-2313-467X]{Diana Dragomir}
\affiliation{Department of Physics and Astronomy, The University of New Mexico, 210 Yale Blvd NE, Albuquerque, NM 87106, USA}
\email{dragomir@unm.edu}

\author[orcid=0000-0002-2482-0180]{Zahra~Essack}
\affiliation{Department of Physics and Astronomy, The University of New Mexico, 210 Yale Blvd NE, Albuquerque, NM 87106, USA}
\email{zessack@unm.edu}

\author[orcid=0000-0003-1361-985X]{Juliana Garc\'ia-Mej\'ia}
\altaffiliation{51 Pegasi B Fellow, MIT Pappalardo Physics Fellow}
\affiliation{Kavli Institute for Astrophysics and Space Research, Massachusetts Institute of Technology, Cambridge, MA 02139, USA}
\affiliation{Center for Astrophysics \textbar\ Harvard \& Smithsonian, 60 Garden Street, Cambridge, MA 02138, USA}
\email{jgarciam@mit.edu}

\author[orcid=0000-0003-3250-2876]{Yang Huang}
\affiliation{School of Astronomy and Space Science, University of Chinese Academy of Sciences, Beijing 100049, People's Republic of China}
\affiliation{National Astronomical Observatories, Chinese Academy of Sciences, Beijing 100101, People's Republic of China}
\email{yanghuang@pku.edu.cn}

\author[orcid=0000-0002-7420-6744]{Jinzhong Liu}
\affiliation{Xinjiang Astronomical Observatory, Chinese Academy of Sciences, Urumqi 830011, People’s Republic of China}
\email{liujinzh@xao.ac.cn}

\author[orcid=0000-0002-9312-0073]{Christopher R. Mann}
\affiliation{National Research Council Canada, Herzberg Astronomy \& Astrophysics Research Centre, 5071 West Saanich Road, Victoria, BC V9E 2E7, Canada}
\email{Christopher.Mann@nrc-cnrc.gc.ca}

\author[orcid=0000-0002-4047-4724]{Hugh P. Osborn}
\affiliation{Physikalisches Institut, Universität Bern, Gesellschaftsstrasse 6, 3012 Bern, Switzerland}
\email{hugh.osborn@unibe.ch}

\author[orcid=0000-0001-8993-5053]{Aleks Scholz}
\affiliation{SUPA, School of Physics \& Astronomy, University of St Andrews, North Haugh, St Andrews, KY16 9SS, United Kingdom}
\email{as110@st-andrews.ac.uk}

\author[orcid=0000-0003-4526-3747]{Andreas Seifahrt}
\affiliation{Gemini Observatory/NSF NOIRLab, 670 N. A'ohoku Place, Hilo, HI 96720, USA}
\email{andreas.seifahrt@noirlab.edu}

\author[orcid=0000-0003-2171-5083]{Patrick Tamburo}
\affiliation{Center for Astrophysics \textbar\ Harvard \& Smithsonian, 60 Garden Street, Cambridge, MA 02138, USA}
\email{patrick.tamburo@cfa.harvard.edu}

\author[orcid=0009-0004-3360-3702]{Shuming Wang}
\affiliation{Xingming Observatory, Urumqi 830011, Xinjiang, People’s Republic of China}
\email{2971749136@qq.com}

\author[orcid=0000-0001-8677-2391]{Shahidin Yaqup}
\affiliation{Xinjiang Astronomical Observatory, Chinese Academy of Sciences, Urumqi 830011, People’s Republic of China}
\email{xiayiding@xao.ac.cn}

\correspondingauthor{T.~Hu, S.~X.~Wang}

\begin{abstract}
TOI-4600\,b and c, originally identified by the Transiting Exoplanet Survey Satellite (TESS) and reported by \cite{Mireles2023}, are a rare pair of transiting long-period giant planets ($\rm P_b=82.7$ days, $\rm P_c=482.8$ days) orbiting an early K dwarf. In this work, we refine the orbital parameters of the TOI-4600 system by combining new TESS photometry, ground-based transit follow-up, and radial velocity (RV) observations from MAROON-X. We obtain improved constraints on planetary masses and eccentricities, and update other parameters, such as the stellar age. For TOI-4600\,b, we measure a mass of $M_p = 74.7^{+4.7}_{-4.4}\,M_{\oplus}$ and an eccentricity of $e=0.153^{+0.020}_{-0.018}$, and $M_p = 212.53^{+13.26}_{-13.03}\,M_{\oplus}$ and $e=0.219^{+0.015}_{-0.018}$ for TOI-4600\,c. We find significant transit timing variations (TTV) in both planets, with semi-amplitudes of approximately $1$\,hr. We derive Transit Spectroscopy Metric values of 16.87 for TOI-4600\,b and 10.09 for TOI-4600\,c, indicating that both planets are promising JWST targets for studying the atmospheres of temperate and cold Jupiters, a relatively poorly characterized sample thus far. These updated parameters and TTV ephemerides are important for planning and interpreting future photometric, spectroscopic, and dynamical studies of the TOI-4600 system.
\end{abstract}



\section{Introduction}\label{sec:intro}

Long‑period transiting exoplanets are intrinsically rare and observationally valuable, owing to their low geometric transit probability and the limited number of observable transits within typical survey baselines \citep{winn2014transitsoccultations, Foreman_Mackey_2016, Herman_2019}. Among them are warm and cold Jupiters, giant planets with orbital periods of tens to hundreds of days. They occupy a particularly important regime, bridging the gap between short-period hot Jupiters and the colder giant planets detected primarily by radial velocity surveys. As they are less subject to stellar tides and irradiation compared with hot Jupiters and often retain non‑circular orbits \citep{Bonomo_2017}, warm and cold Jupiters provide key insights into giant planet formation, migration, and dynamical evolution \citep{Chatterjee2008, Dawson_2013}.


In particular, warm and cold Jupiters provide favorable conditions for atmospheric characterization at relatively low temperatures, where nitrogen chemistry becomes accessible. In this regime, ammonia (NH$_3$) is expected to be a major nitrogen-bearing species and can be detected in planetary atmospheres, offering a powerful tracer of giant-planet formation locations \citep[e.g.,][]{Fortney2020}. Nitrogen-bearing and oxygen-bearing species abundances depend sensitively on disk temperature and radial composition gradients; therefore, measurements of NH$_3$ can provide independent constraints on planetary accretion histories and formation pathways \citep{Cridland2020}. In contrast, most atmospheric studies to date have focused on hot Jupiters, for which carbon-bearing species dominate the observable chemistry, and formation inferences based primarily on carbon diagnostics, such as C/O ratios \citep{Oberg_2011,Madhusudhan_2014,Brewer2016}, are subject to significant degeneracies when used alone \citep{Turrini_2021}. Extending atmospheric measurements to cooler giant planets is therefore essential for breaking these degeneracies and for providing more robust constraints on giant-planet formation theories \citep{Ohno2023_1,Ohno2023_2}.


The diagnostic power of warm and cold Jupiters is further enhanced in systems hosting multiple transiting planets. The simultaneous transit of multiple planets implies a low mutual inclination among the planets \citep{Fabrycky_2014}. Such coplanarity strongly disfavors violent dynamical processes, such as planet–planet scattering, which generically excite large mutual inclinations and would likely disrupt the transiting configuration \citep{Chatterjee2008}. For long-period giant planets, the combination of weak tidal interactions and dynamically cold, coplanar multi-planet architectures implies that the observed stellar obliquity has experienced minimal post-formation evolution. Consequently, obliquity measurements in multi-transiting warm and cold Jupiter systems are expected to closely trace the primordial misalignment of the proto-planetary disk \citep{batygin2013primordialoriginmisalignmentsstellar, Lai_2014}, a potentially common architecture that contrasts with our own solar system \citep{Biddle2025}.




TOI-4600 b and c are a pair of transiting, long-period giant planets ($\rm P_b=82.7$ days, $\rm P_c=482.8$ days) confirmed by \cite{Mireles2023}. As the only known pair of transiting giants straddling the water snow line, this system provides a unique testbed for investigating how radial composition gradients within the protoplanetary disk influence final planetary compositions. Atmospheric observations on oxygen tracers in both planets would provide a direct, internally consistent test of whether their current atmospheres reflect their formation locations relative to snow lines \citep[e.g.,][]{Schneider2021_a, Schneider2021_b}, or if this imprint was erased by other processes during or after their formation \citep[e.g.,][]{Molliere2022}. In addition, the temperate regime of these planets allows for further formation constraints using simultaneous C, N, and O measurements. This wealth of information makes TOI-4600 a potential benchmark system for studying exoplanet formation and evolution.

In this work, we present a revised characterization of the TOI-4600 system. Using new space- and ground-based photometry, we identify transit timing variations (TTVs) for both planets b and c. By combining these photometric data with ground-based radial velocity (RV) measurements, we determine the masses and eccentricities of both planets and improve the transit time predictions for future events. In Section~\ref{section:2}, we present the photometric and spectroscopic data sets used in our analysis. In Section~\ref{section3}, we determine the stellar properties of TOI-4600, including its age and projected rotational velocity. In Section~\ref{section4}, we present the results of our transit modeling, radial velocity analysis, and photodynamical fits, and derive the planetary and orbital parameters of TOI-4600\,b and c. In Section~\ref{section5}, we discuss the TOI-4600 system in the context of other multi-transiting giant-planet systems, investigate the physical origin of its large-amplitude transit timing variations, and outline prospects for further characterization. Finally, in Section~\ref{section6}, we summarize our main conclusions and outline avenues for future work.

\section{Observations\label{section:2}}

\subsection{TESS photometry\label{sec:tess}}
The TESS mission is an all-sky survey that aims to discover transiting exoplanets \citep{ricker2015}. The mission has a field of view of $24^{\circ}\times 96^{\circ}$ and observes the sky in sectors, with each sector lasting approximately 27 days. 

TOI-4600 is in the continuous viewing zone of TESS and up till November 1, 2025, TESS has observed TOI-4600 (TIC 232608943) with Cameras 3 and 4 for a total of 38 sectors. Sector 86 in December 2024 is the newest sector and the next time TESS observes TOI-4600 will be sector 117 in May 2027. During the first 12 sectors (Sectors 14--19 and 21--26), TOI-4600 was only observed in 30-minute cadence Full Frame Images (FFIs). In the latter 26 sectors (Sectors 40, 41, 47--49, 51--60, 74--76, 78--80, and 82--86), the mission collected 2-minute cadence data for TOI-4600. Among them, sectors 16,22,25,47,53,56,59,80,83,86 covered 10 transits of planet b, while sectors 17 and 53 covered two transits of planet c. We only used these sectors that contained the transits of planet b and c for all analyses in this paper. 

The TESS Science Processing Operations Center (SPOC; \citealt{Jenkins2016}) 
pipeline at NASA Ames Research Center calibrates the 30-minute cadence FFI data. Then the TESS-SPOC pipeline \citep{Caldwell2020} extracts photometry from these SPOC-calibrated FFIs. 

The SPOC pipeline also analyzes the 2‑minute cadence data and, for each sector, delivers two distinct light-curve products: Simple Aperture Photometry (SAP) and Presearch Data Conditioning Simple Aperture Photometry (PDCSAP; \citealt{Smith2012, Stumpe2012, Stumpe2014}). The PDCSAP light curves incorporate corrections for instrumental systematics, remove outliers, and account for crowding effects and are used in our analysis.

\begin{figure*}[htbp]
    \centering

    \begin{subfigure}[b]{\textwidth}
        \centering
        \includegraphics[width=\textwidth]{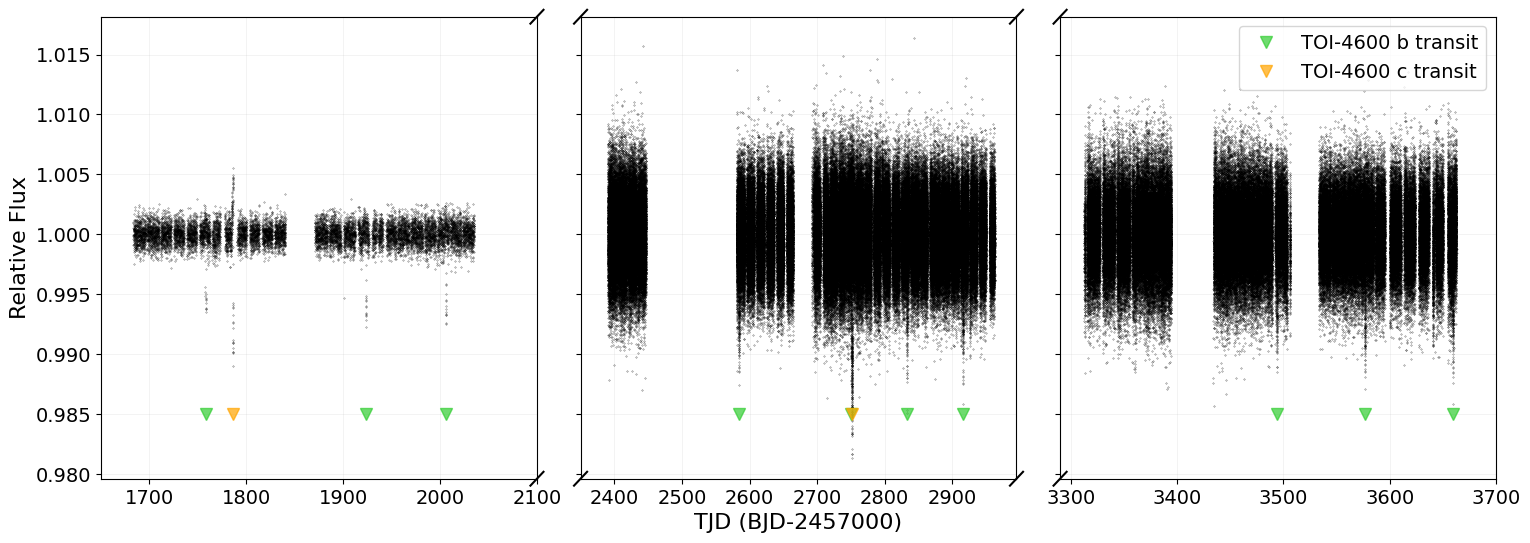}
        
        \label{fig:full_lightcurve_a}
    \end{subfigure}

    \vspace{1em} 

    \begin{subfigure}[b]{\textwidth}
        \centering
        \includegraphics[width=\textwidth]{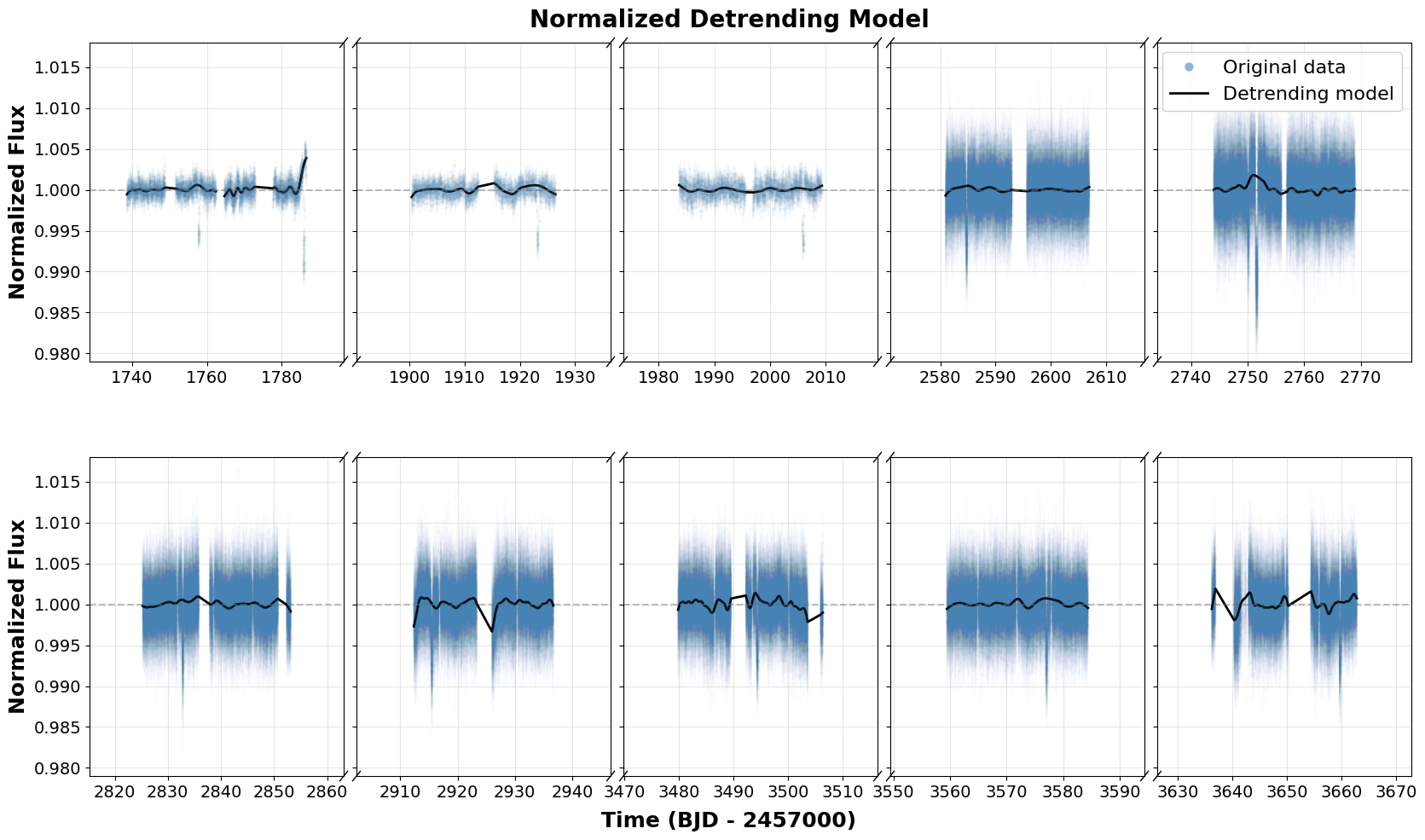}
        
        \label{fig:full_lightcurve_b}
    \end{subfigure}

    \caption{Top: TESS light curves of TOI‑4600. Green triangles indicate the transits of TOI‑4600 b, while orange triangles label the transits of TOI‑4600 c. The left-most panel corresponds to the 30‑minute cadence data from sector 14--19 and sector 21--26. The middle panel shows the 2‑minute cadence data from sector 40--41, sector 47--49, and sector 51--60. The right panel shows the 2‑minute cadence data from sector 74--76, sector 78--80, and sector 82--86. Bottom: Detrending model of the lightcurves from sectors that possess transit events. The black line represents the detrending model using GP, while the blue dots represent the photometric data. Both the model and the data are first normalized by the median of their respective sectors.}
    \label{fig:full_lightcurve_combined}
\end{figure*}

 We queried Mikulski Archive for Space Telescopes (MAST)\footnote{\url{https://archive.stsci.edu}} programmatically with the \texttt{Lightkurve} package \citep{Lightkurve} to download the corresponding SPOC light-curve and DV products for all relevant TESS sectors. We then detrended the TESS light curve by fitting a Gaussian Process (GP) model with the \texttt{Juliet} package \citep{Espinoza2019} after masking the in-transit data. In this framework, \texttt{Juliet} constructs the GP model using the \texttt{Celerite} package \citep{Foreman-Mackey_2017} with a Mat\'ern 3/2 kernel. Figure~\ref{fig:full_lightcurve_combined}.a presents the original light curves while Figure~\ref{fig:full_lightcurve_combined}.b presents the best-fit GP models. All TESS data products used in this work are available via MAST at doi:\href{https://doi.org/10.17909/kr6d-vk05}{10.17909/kr6d-vk05}.
\subsection{Ground-based photometry}
\label{observatory}

To ensure reliable transit timing measurements, we only retain light curves that fully cover at least one complete ingress or egress.

\subsubsection{The \upshape{St Andrews} \textit{Observatory}} \label{sec:standrews}

We observed the egress of TOI-4600c on 2023 October 17 using the James Gregory Telescope (JGT) at the University of St Andrews, UK. The JGT is a 37-inch Schmidt--Cassegrain telescope equipped with a CCD camera providing a $50 \times 40$ arcmin field of view at prime focus. The telescope has been used extensively for the follow-up of variable stars, including exoplanet transits (e.g., Bozhinova et al.~2016; Hay et al.~2016; Scholz et al.~2021).

A total of 210 images were obtained in four sequences of 50 exposures each, plus an additional 10 exposures, with an exposure time of 60~s per frame. The observations were conducted between 01:00 and 06:00~GMT. Although the sky remained clear throughout the night, the target was observed at relatively high airmass ($\sim$1.7--2.0) due to its position and the timing of the transit. The observing sequence was terminated when the telescope reached its hour angle limit.

The data were reduced using standard bias frames and sky flats obtained during morning twilight on the same night. The resulting light curve clearly shows the egress of the transit.

\subsubsection{The \upshape{Tierras} \textit{Observatory}} \label{sec:tierras}
We observed the egress of TOI-4600~c with the 1.3-m \textit{Tierras} Observatory \citep{GarciaMejia2020} located atop Mt. Hopkins in southern Arizona on UT~2023~October~17. We obtained 152 images at a continuous cadence with an exposure time of 30 seconds, covering a $\text{BJD}_{\text{TDB}}$ range of $2460234.57446$ to $2460234.64251$. All images were taken in the custom \textit{Tierras} filter, which is centered at $863.5$~nm with a FWHM of $40.2$~nm. These observations, which occurred during an instrument maintenance run, were mistakenly performed with auxiliary tracking switched off, resulting in PSF elongation of about 10 pixels along the R.A. axis in each exposure. We performed circular aperture photometry on TOI-4600 and 13 reference stars in the images with the \textit{Tierras} photometric pipeline \citep{Tamburo2025}, finding that a 13-pixel radius minimized the light curve scatter for TOI-4600. We note that we tested elliptical apertures to better match the elongated PSF shapes but found no significant improvement over circular apertures, which is not surprising given the brightness of TOI-4600.

The \textit{Tierras} observations revealed a significant rise and then plateau of the flux from TOI-4600 across the observations, consistent with the egress of TOI-4600~c. Since a similar structure was not observed in any of the reference star light curves, we concluded that the observations did detect the transit egress. 

\subsubsection{The \upshape{Kourovka} \textit{Observatory}} \label{sec:kourovka}
We observed the ingress of TOI-4600~c with the 2x0.4-m \textit{Kourovka} Observatory of Ural Federal University \citep{Lipunov2010}, located in Sverdlovsk Oblast, 80 km~west of Yekaterinburg, Russia, on UT~2023~October~16. We obtained 82 images at a continuous cadence in V and R filters with an exposure time of 180 seconds, covering a $\text{BJD}_{\text{TDB}}$ range of $2460234.14826$ to $2460234.41538$. Pixel scale = 1.85~arcsec/pixel, mean FWHM of PSF is 4.1~arcsec for filter V and 4.5~arcsec for filter R. Photometric aperture radius is 3~pixels or 5.5~arcsec for filter V and 5.6~arcsec for filter R.

\subsubsection{The Nanshan Station of the Xinjiang Astronomical Observatory}
We observed TOI-4600 c on UT 2025 February 10 using the Photometric Auxiliary Telescope-17 (PAT-17, diameter 43 cm) at the Nanshan station of the Xinjiang Astronomical Observatory (XAO; \citealt{2025MNRAS.537.2258L, 2025RAA....25j5003W}). The telescope is equipped with a QHY4040 sCMOS camera ($4118 \times 4136$ pixels) and a Sloan filter system. We obtained time-series photometry in the Sloan $r$ band with an exposure time of 120 s, covering a $\mathrm{BJD_{TDB}}$ range of 2460717.252185 to 2460717.553791. The camera has a pixel scale of $0.63\arcsec\ \mathrm{pixel}^{-1}$ and a field of view of $43\arcmin \times 43\arcmin$.
\subsection{Space-based Photometry}

We observed TOI-4600~c semi-continuously with the Near-Earth Object Surveillance Satellite (NEOSSat; \citealp{laurin2008neossat}) over 1.9 days, from 2023-10-15 23:15 (2460233.468937~BJD) to 2023-10-17 21:05 (2460235.378672~BJD). Between interruptions for Earth-eclipse events and necessary spacecraft operations, we achieved an on-target duty cycle of $\sim33$\%. We observed with an exposure time of 40~seconds, and a 512$\times$512 pixel (25$\times$25') image sub-raster. This allowed for the inclusion of sufficient reference stars in the frame to perform a principal component analysis (PCA) normalization routine to remove systematics from the photometric time series. We removed 64 data points between $\mathrm{BJD}=2460234.65$ and $2460234.70$ because they exhibited significant systematic errors. The final light curve had a point-to-point scatter of $\sim$8.9~ppt, which binned to 7.3~ppt in 10-minute bins. We made a clear detection of the 11~ppt event.

\subsection{MAROON-X Spectroscopy}
\label{sec:MAROON-X}

We obtained high-resolution optical spectra with the MAROON-X spectrograph \citep{Seifahrt2018}, installed on the 8.1\,m Gemini North Telescope on Maunakea, Hawaii. MAROON-X operates at a resolving power of approximately \(\lambda / \Delta \lambda \approx 85{,}000\) and covers the 500--920\,nm wavelength range. The instrument employs a fiber-fed, bench-mounted design and achieves exceptional thermo-mechanical stability, which allows us to reach sub-meter-per-second RV precision.

MAROON-X employs two CCD detectors that capture spectra over separate wavelength intervals: a blue arm spanning 500–670\,nm and a red arm spanning 650–920\,nm. Both arms acquire data simultaneously during each observation. Our dataset comprises 22 RV measurements from the blue arm and 22 from the red arm, obtained at matching epochs between UT 2024-02-28 and UT 2024-07-30 under Program ID GN-2024A-Q-304. The exposure time was 1200\,s for the first epoch and 900\,s for each of the subsequent 21 epochs in both arms.

We processed the MAROON-X spectra with a Python3 implementation of the \texttt{SERVAL} pipeline \citep{Zechmeister2018}, rewritten and adapted for MAROON-X. We first co-add all available spectra to construct a high signal-to-noise ratio (SNR) master template for the target, and then we derive relative RVs by matching this template to each individual spectrum. 

The red-arm spectra reach a median SNR of 68 per resolution element, with a mean SNR of 69.3, a range of 26--100, and a standard deviation of 21.0. The blue-arm spectra reach a median SNR of 54, with a mean SNR of 55.0, a range of 22--78, and a standard deviation of 15.7. All 22 exposures in each arm meet our quality criteria, so we include the full set of 44 RV measurements in the analysis. We list the final MAROON-X RVs and their uncertainties in Table~\ref{tab:maroonx_rv_combined}.

\subsection{TRES Spectra}

We collected a set of nine reconnaissance spectra of TOI-4600 with the Tillinghast Reflector Echelle Spectrograph (TRES) mounted on the 1.5\,m Tillinghast telescope at the Fred L. Whipple Observatory (FLWO) on Mt.\ Hopkins, Arizona \citep{tres}. The observations were acquired between UT 2021 May 31 and UT 2022 September 28, covering the complete orbit of TOI-4600 b and the orbital phases $0$ to $0.5$ of TOI-4600 c. The signal-to-noise ratios of the spectra range from 21.9 to 35.4. Each spectrum was extracted following the standard procedures described by \cite{Buchhave_2010}. Multi-order relative radial velocities were then measured via order-by-order cross-correlation, using the strongest observed spectrum as a template.
We used these TRES data to estimate the chromospheric activity level of the host star and to impose additional constraints on the stellar age.

\section{Stellar properties} \label{section3}


\subsection{Spectral analysis} \label{sec:spectral_analysis}

We analyzed the MAROON-X spectra using \texttt{SpecMatch-Emp} \citep{Yee2017_specmatch} to constrain the stellar effective temperature ($\rm T_{eff}$) and metallicity ($\rm [Fe/H]$), providing priors for subsequent isochrone fitting. Briefly, \texttt{SpecMatch-Emp} compares the input
spectrum with a library of optical spectra observed with Keck/HIRES by the California Planet Search. It interpolates the five best-matched spectra and returns stellar parameters and uncertainties To align with the wavelength range of \texttt{SpecMatch-Emp}, the final MAROON-X spectrum used for the fit spanned from $\sim$4994 \AA\ to $\sim$6294 \AA. 

We began with the stellar template generated by \texttt{SERVAL}. To normalize the spectra of each order, we selected the highest point within each 4 \AA\ windows, and then used a cubic spline to connect these points to build the continuum for that order. We did not fine-tune this simple normalization further, as \texttt{SpecMatch-Emp} is robust to handle the imperfect normalization. We then used \texttt{SpecMatch-Emp} to merge all orders into a unified spectrum, shift the spectrum to the reference frame of the \texttt{SpecMatch-Emp} library, and perform the match to derive the stellar parameters. As a result, the derived $\rm T_{eff}$ is $5040 \pm 110$ K and the $\rm [Fe/H]$ is $0.20 \pm 0.09$ dex.

\subsection{SED and isochrone fitting} \label{sec:isochrone}

We constrained the broadband spectral energy distribution (SED) and host-star evolutionary properties using our pipeline \texttt{SEDEX\_ISO}\footnote{\url{https://github.com/Jiayue-Zhang99/SEDEX_ISO.git}}, which extends SEDEX \citep{Yu2023_SEDEX} by coupling SED fitting to isochrone-based inference. Multiband photometry was assembled automatically from archival catalogs, including Gaia DR3 \citep{Riello2021_GaiaPhot} and a set of commonly used optical and infrared surveys (e.g., APASS DR9 \citep{Henden2016_APASS}, Sloan Digital Sky Survey (SDSS) DR13 \citep{Alam2015_SDSS}, Pan-STARRS DR1 \citep{chambers2019panstarrs1surveys}, 2MASS \citep{Cutri2003_2MASS}, and ALLWISE \citep{Cutri2021_ALLWISE}) supported by the code configuration.

For the SED analysis, we fitted the broadband photometry with high-resolution synthetic spectra from the MARCS library \citep{Gustafsson2008}. We adopted the spectroscopic effective temperature and metallicity derived in Section~\ref{sec:spectral_analysis}, together with a surface gravity prior based on the Gaia DR3 catalog ($\log g = 4.527$; \citealt{GaiaDR3}), as an external constraint in the fit. We also incorporated the Gaia-based distance constraint in the inference. The SED fitting then provided posterior estimates of the bolometric flux scale, angular radius, broadband extinction (when applicable), and a self-consistent set of atmospheric parameters, including $T_{\rm eff}$, $\log g$, and metallicity $\rm [Fe/H]$.

These SED-derived atmospheric parameters were subsequently propagated to the isochrone analysis, which was carried out with the \texttt{isochrones} package \citep{Morton2015} using MIST model grids \citep{Dotter2016_MIST1, Choi2016_MIST2}. The isochrone fitting yielded posterior estimates of the stellar mass, radius, mean density, luminosity, and age. For the SED fitting, we report posterior-weighted mean values and weighted standard deviations. For the isochrone fitting, we report the mean and standard deviation of the posterior samples. The adopted stellar parameters are summarized in Table~\ref{tab:stellar_parameter}. These parameters are all consistent with those reported in \cite{Mireles2023}.

\begin{deluxetable*}{lll}
\tablewidth{1.0\textwidth}
\tablecaption{Adopted stellar parameters for TOI-4600. Reported values are given as the mean $\pm$ 1$\sigma$ uncertainty. \label{tab:stellar_parameter}}
\tablehead{\colhead{Parameter} & \colhead{Value} & \colhead{Source}  } 
\startdata
Stellar Mass (M$_\odot$)       & $0.880\pm0.048$ & Isochrone fitting\\ 
Stellar Radius (R$_\odot$)     & $0.817\pm0.035$ & Isochrone fitting\\ 
Effective Temperature (K)      & $5065\pm101$    & SED fitting\\ 
$\log g_*$ (cgs)               & $4.53\pm0.20$   & SED fitting\\ 
$\rm [Fe/H]$ (dex)             & $0.20\pm0.09$   & SED fitting\\ 
Stellar Density (g/cm$^3$)     & $2.27\pm0.32$   & Isochrone fitting\\ 
Stellar Luminosity (L$_\odot$) & $0.409\pm0.015$ & Isochrone fitting\\ 
Age (Gyr)                      & $3.9\pm2.2$     & Isochrone fitting\\ 
                               & $5.24^{+1.53}_{-2.10}$ &Kinematics\\
$v_* \sin i$ (km/s)            & $\lesssim 2$    & Section \ref{sec:vsini} \\ 
\enddata
\end{deluxetable*}

\subsection{Stellar age}


We estimated the chromospheric activity level of the host star TOI-4600 using the Ca\,\textsc{ii} H \& K lines, following the classical Mount Wilson Observatory (MWO) definition of the $S$-index. The analysis was performed on the TRES spectrum, and MAROON-X spectra do not cover the Ca\,\textsc{ii} H \& K lines. The original spectrum has a resolving power of $R \sim 48{,}000$, which we degraded to $R \sim 30{,}000$ in order to approximate the spectral resolution of the MWO system. The flux was integrated within two triangular bandpasses centered at 3934.78~\AA\ (Ca\,\textsc{ii} K) and 3969.59~\AA\ (Ca\,\textsc{ii} H), each with a full width of 1.09~\AA. The continuum level was estimated using 10~\AA-wide reference bandpasses located on 3901~\AA\ and 4001~\AA\ . Following \citet{1984ApJ...279..763N}, the $S$-index is defined as
\begin{equation}
S = \alpha \frac{N_H + N_K}{N_R + N_V},
\end{equation}
where $N_H$ and $N_K$ are the fluxes measured in the triangular H and K bandpasses, $N_R$ and $N_V$ are the fluxes in the two reference continuum bands, and $\alpha$ is the calibration constant. Using this relation with $\alpha \approx 1$ \citep{2025ApJ...984....2H}, we measured $S$-index $\approx 0.029$. Although this value has not been fully calibrated onto the Mount Wilson scale, inspection of the spectrum shows no clear emission reversal in the cores of the Ca\,\textsc{ii} H \& K lines, indicating weak chromospheric activity of TOI-4600. This suggests that TOI-4600 is not a young, magnetically active star. The inferred chromospheric emission is lower than that observed in young ($\sim 1$ Gry), rapidly rotating stars, and is broadly consistent with an intermediate to relatively old stellar age. 

We further estimate the kinematic age of TOI-4600 following the method proposed by \citet{2024ApJ...966L..27C}. The age is inferred by mapping its kinematic properties onto those of a reference sample with well-determined ages. We select thin-disk (low-$\alpha$) stars within 1 kpc from the subgiant sample of \citet{2025NatAs...9..101X}, which provides precise ages derived from isochrone fitting for evolved subgiant stars.
The selected stars are binned in the $V$ and $\sqrt{U^2 + V^2}$ space, using a bin width of 5 km s$^{-1}$ in both dimensions. For TOI-4600, we identify the corresponding kinematic bin and define its age as the median age of stars within that bin. The associated uncertainty is taken as the age dispersion of stars in the same bin.
We obtain a kinematic age of $\tau_{\rm TOI-4600} = 5.24^{+1.53}_{-2.10}$\,Gyr, which is consistent with its low level of stellar activity and our isochrone-fitting results.




\section{Analysis and Results\label{section4}} 

While \cite{Mireles2023} measured the orbital periods of TOI-4600 b and c, the masses of both planets remained poorly constrained due to limited RV data. With the additional RV measurements, we aim to accurately constrain these masses. Because the RVs do not cover enough phases for planet c, its orbit is poorly constrained when using free priors for eccentricity ($e$) and the argument of periastron ($\omega$). However, we found the RV orbit can be better constrained when adding priors from transits. Since we now have 18 more sectors of TESS data than \cite{Mireles2023}, we first performed a transit fit to establish more precise priors for the subsequent RV modeling.

\subsection{Transit and TTV analysis with \texttt{juliet}}

We performed the transit analysis using the Python package \texttt{juliet}. We first fitted the detrended light curves of TOI-4600 b and c separately. The transit model is generated using \texttt{batman} and includes corrections for instrument-dependent systematics. Limb darkening is described using a quadratic law \citep{MA2002}, parameterized by $q_0$ and $q_1$ following \cite{Kipping2013_ld}. 

We adopted a Gaussian prior on the stellar density ($\rho_\star$) based on our stellar characterization, while the parameters $r_1$, $r_2$, $\sqrt{e}\sin\omega$, $\sqrt{e}\cos\omega$, $q_0$, and $q_1$ were assigned uniform priors. Here we adopt the $(r_1, r_2)$ parametrization of \cite{Espinoza_2018}, where $r_1$ maps to the impact parameter and $r_2$ to the planet-to-star radius ratio. Posterior sampling was performed using the nested sampler \texttt{dynesty} \citep{2020MNRAS.493.3132S}.

We found that the transit times inferred from the ground-based observations showed significant deviations from the linear ephemeris derived from the global fit, suggesting the presence of transit timing variations (TTVs) in the system.

To further investigate this, we first analyzed TTVs without assuming a dynamical model. For TOI-4600 b, we allowed the mid-transit time of each transit to vary independently with uniform priors and refitted the data using \texttt{juliet}. The resulting individual transit times were then used to derive a reference orbital period $P$ and transit epoch $T_0$ via a linear least-squares fit. The observed transit times deviate from this linear ephemeris, confirming the presence of TTVs. The TTVs are computed as the differences between the observed mid-transit times and the predicted values from this reference model.

For TOI-4600 c, we analyzed transit observations obtained across multiple epochs and facilities. The 0th and 2nd transits were observed by \textit{TESS}, while the 4th transit was observed by \textit{Nanshan Observatory}. The 3rd transit includes observations from NEOSSat, St.~Andrews, Bell Burnell, and additional publicly available datasets from ExoFOP. The selection of ground-based datasets used in this analysis is described in Section~\ref{observatory}.

 We first fitted the combined \textit{TESS} photometry from the 0th and 2nd transits to constrain the transit shape parameters. We then jointly fitted these data with the selected ground-based and NEOSSat observations of the 3rd transit to determine its mid-transit time. For the 4th transit, we applied a similar joint fitting procedure using the \textit{TESS} photometry together with the Nanshan observations. The individual non-\textit{TESS} light curves used for the 3rd- and 4th-transit timing analyses, together with their best-fit transit models, are shown in Figure~\ref{fig:toi4600c_ground_lcs}.

 \begin{figure*}[htbp]
    \centering
    \includegraphics[width=\textwidth]{ground.png}
    \caption{
    Multi-instrument photometry used to determine the mid-transit times of the 3rd and 4th observed transits of TOI-4600\,c. 
    Each panel shows the detrended light curve from an individual non-\textit{TESS} facility. 
    Black points show the observed relative fluxes with 1$\sigma$ uncertainties, and red curves show the best-fit transit models. 
    The horizontal axis gives the time relative to the fitted mid-transit time of each transit, $T-T_c$, in hours. 
    }
    \label{fig:toi4600c_ground_lcs}
\end{figure*}

The posterior distributions of $\sqrt{e}\cos\omega$ and $\sqrt{e}\sin\omega$ derived from the fit including the 4th transit are consistent with those obtained from the 0th and 2nd transits, while the 3rd transit shows modest deviations. Given that the retained datasets for the 3rd transit provide sufficient phase coverage, we consider the inferred mid-transit time to be reliable.

Using the mid-transit times of the 0th, 2nd, 3rd, and 4th transits, we derived a linear ephemeris $(T_0, P)$ via least-squares fitting. Similar to planet b, the transit times of TOI-4600 c show clear deviations from this linear model, indicating the presence of TTVs. The TTVs are computed relative to this reference ephemeris.

\begin{figure*}[htbp]
    \centering
    \begin{subfigure}[b]{0.48\textwidth}
        \centering
        \includegraphics[width=\textwidth]{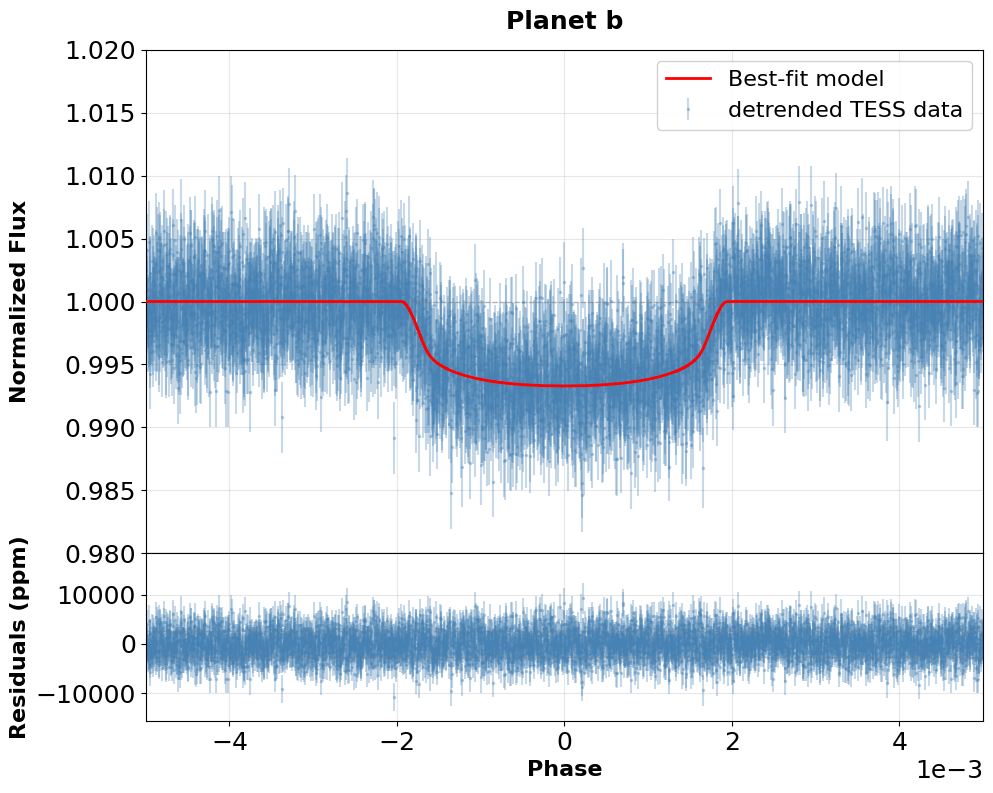}
    
        \label{phase_b}
    \end{subfigure}
    \hfill
    \begin{subfigure}[b]{0.48\textwidth}
        \centering
        \includegraphics[width=\textwidth]{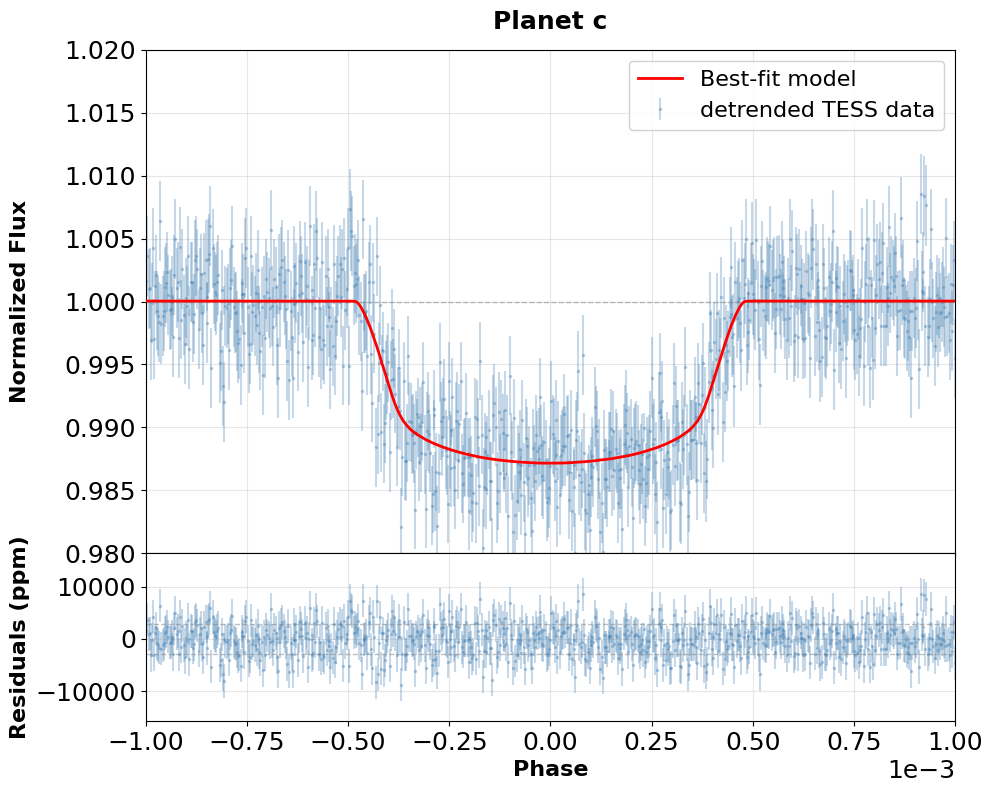}
    
        \label{phase_c}
    \end{subfigure}
    \caption{The left panel shows phase-folded data and the best-fit model with residuals of TOI‑4600 b, and the right panel shows those of planet c.}
    \label{phase-folded}
\end{figure*}

The fitted and derived parameters from the transit analysis are reported in Table~\ref{tab:fitted-params-transit}. Phase-folded light curves and best-fit models for both planets are shown in Figure~\ref{phase-folded}.

\subsection{RV analysis with \texttt{juliet}}

We fitted the MAROON-X radial velocities using the \texttt{juliet} package, which employs \texttt{radvel} \citep{Fulton2018} to generate Keplerian RV models. Bayesian inference was performed with the nested sampler \texttt{dynesty} \citep{Speagle2020}.

In the RV model, each planetary orbit was parameterized by the orbital period $P$, eccentricity $e$, argument of periastron $\omega$, RV semi-amplitude $K$, and time of inferior conjunction $T_{\rm conj}$. In practice, rather than fitting $e$ and $\omega$ directly, we adopted the parametrization $\sqrt{e}\cos\omega$ and $\sqrt{e}\sin\omega$, which improves sampling efficiency and reduces biases near circular orbits. The orbital period $P$ and time of inferior conjunction $T_{\rm conj}$ were fixed to the values derived from the transit analysis.

We included independent systemic velocity offsets, $\mu_{\rm red}$ and $\mu_{\rm blue}$, for the red- and blue-arm RV datasets, respectively, to account for possible zero-point differences between the two arms. We also fitted independent white-noise jitter terms, $\sigma_{w,\rm red}$ and $\sigma_{w,\rm blue}$, for the two arms.

\begin{figure*}[htbp]
    \centering

    \begin{subfigure}[b]{\textwidth}
        \centering
        \includegraphics[width=\textwidth]{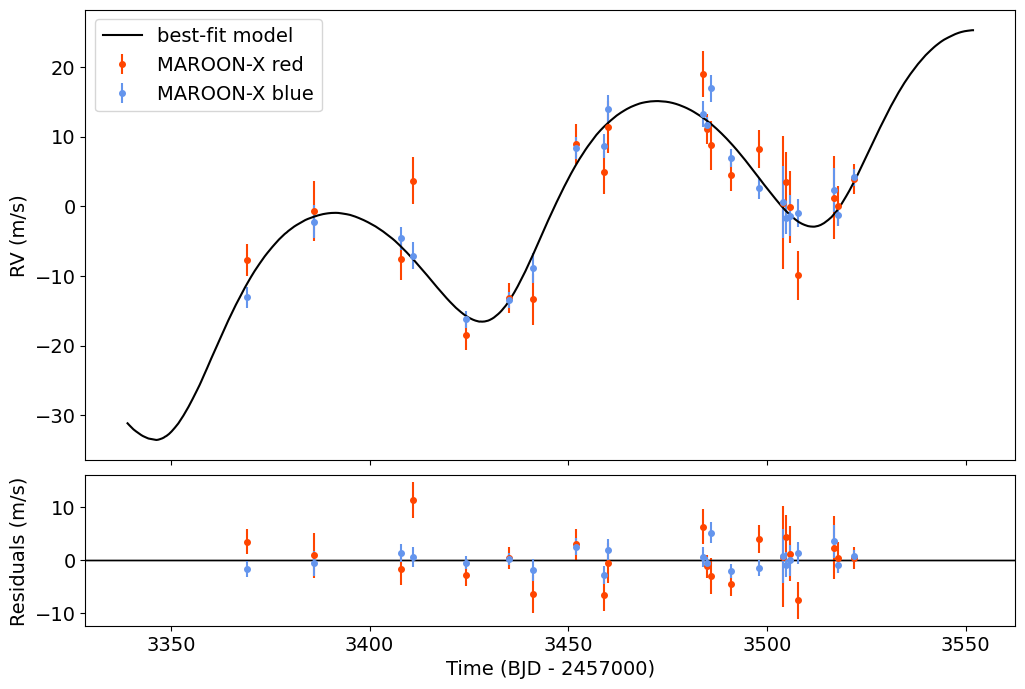}
        \label{fig:rv_model}
    \end{subfigure}

    \vspace{1em}

    \begin{subfigure}[b]{\textwidth}
        \centering
        \includegraphics[width=\textwidth]{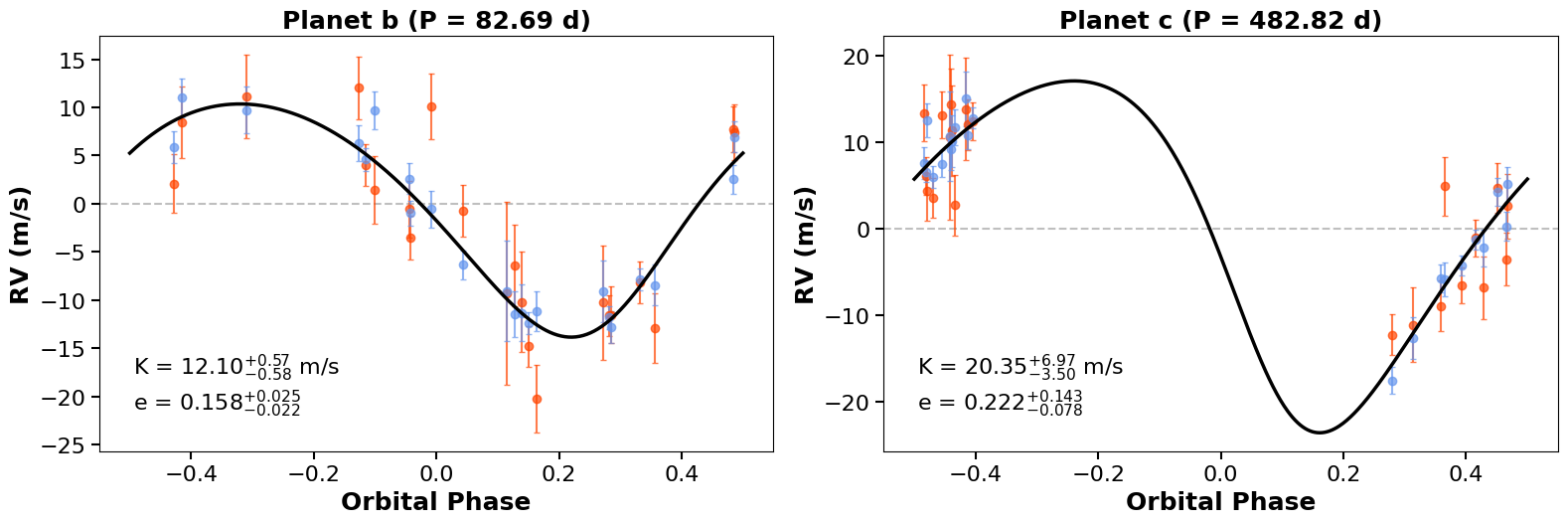}
        \label{fig:rv_phase}
    \end{subfigure}

    \caption{
    MAROON-X radial velocities and best-fitting two-planet Keplerian model. 
    The upper panel shows the measured RVs as a function of time (BJD $-$ 2457000) for the red and blue arms, overplotted with the best-fitting model. 
    The middle panel shows the residuals after subtracting the model from the data. 
    The lower panels show the phase-folded RV signals and best-fitting models for TOI-4600 b and TOI-4600 c, after subtracting the contribution from the other planet.
    }
    \label{fig:rv_overall}
\end{figure*}

The fitted and derived parameters are listed in Table~\ref{tab:rv-fitted-params}, together with the adopted priors. The full RV time series, the best-fitting two-planet Keplerian model, and the phase-folded RV signals of the individual planets are shown in Figure~\ref{fig:rv_overall}. The fit yielded radial-velocity semi-amplitudes of \(K_b = 12.10^{+0.57}_{-0.58}\,\mathrm{m\,s^{-1}}\) for TOI-4600 b and \(K_c = 20.35^{+6.97}_{-3.50}\,\mathrm{m\,s^{-1}}\) for TOI-4600 c.

\subsection{Photodynamic model}

We modeled the available photometric and RV data for TOI-4600 using a photodynamical model that includes a dynamical component accounting for planetary mutual gravitational interactions and predicting planetary transit light curves. Our model follows the framework of \citep{Wang2024}, as detailed below.

Each planet orbit is initialized with osculating parameters of orbital period $P$, eccentricity $e$, argument of pericenter $\omega$, impact parameter $b$, ascending node $\Omega$, and time of inferior conjunction $T_{\rm conj}$. 
The system is described in Jacobian coordinates, with osculating orbital parameters referenced to BJD = 2458745.0. We integrate the system’s motion using the symplectic \texttt{WHFast} integrator \citep{whfast} within the \texttt{rebound} package \citep{rebound}, considering only Newtonian gravitational interactions. The integration timestep is fixed at 1.5 days, approximately 1/50 of the innermost planet’s orbital period, and is performed in the center-of-mass frame. The $+z$-axis points toward the observer. At each observed epoch of photometric and RV data, we record the barycentric coordinates and velocities of all bodies, using the z-component of stellar velocity to compare with radial velocity measurements. 

Light curves are synthesized from the relative coordinates of different bodies with the quadratic limb-darkening model \citep{MA2002}, implemented in \texttt{batman} \citep{Kreidberg2015}. We ignored light-travel-time effects (LTTEs), as their correction to transit timing ($\lesssim$1.6 minutes) is smaller than the observational uncertainty. The total loss of light is assumed to be the sum of losses due to all planets, without accounting for planet-planet occultations \citep{Masuda2013}. Long-cadence light curves are super-sampled by a factor of 10. As we model the light curves from different instruments with distinct wavelength coverage, we assign independent additive photometric jitter terms, $\sigma_{\rm jitter}$, and quadratic limb-darkening coefficients to each instrument. However, partial transits and low SNR of data acquired from the ground-based facilities make it unlikely to precisely constrain the limb-darkening coefficients. Therefore, we assign Gaussian priors to the Kourovka (R, V), SOFAR (g'), Tierras (863.5~nm, 40.2~nm FWHM), and WST (J) centered on the theoretical values from \texttt{EXOFASTv2} \footnote{https://astroutils.astronomy.osu.edu/exofast/limbdark.shtml}.

Finally, we adopted a Gaussian prior of 0.817$\pm$0.035 R$_\odot$ for stellar radius, and 0.88$\pm$0.05 M$_\odot$ for stellar mass, informed by our stellar characterization (Section \ref{section:2}). The planet orbital eccentricities $e$ are assigned with a prior probability $\sim1/e$ to cancel the implicit prior of $\sim e$ when uniformly sampling the eccentricity vectors ($e\cos\omega$, $e\sin\omega$). The prior bounds are summarized in Table \ref{tab:fitted_parameter}.

The log-likelihood of photometric data, $\mathcal{L_{\rm pho}}$, follows the below expression:

\begin{equation}
\label{eq:photo_lh}
    \log \mathcal{L_{\rm pho}} = -\frac{1}{2}\sum_{i}^{N} \left ( \frac{(f_{\rm i,obs}-f_{\rm i,model})^2}{\sigma_i^2}  + \log2\pi\sigma_i^2 \right )
\end{equation}
where $f_{\rm i,obs}$ is the $i$-th observed flux, $f_{\rm i,model}$ is the flux evaluated by the photodynamical model, $\sigma_i^2 = \sigma_{\rm i,obs}^2 + \sigma_{\rm i,jitter}^2$ is the quadratic sum of observed and additive photometric jitters, and $N$ is the total number of photometry data. 

For the joint photodynamical and RV joint modeling, four more parameters are required, resulting in 49 adjustable parameters for the joint model in total. The likelihood of the joint model is the linear combination of Eq \ref{eq:photo_lh} and RV likelihoods.

We first optimized the dynamical model against the TTV and RV data using \texttt{scipy.optimize.curve\_fit}, assuming a coplanar configuration. After obtaining an initial best-fit solution, we initialized a Gaussian ball of perturbed parameters around the optimized result, relaxed the coplanarity assumption by allowing the inclinations to vary, and sampled the posterior distributions using the affine-invariant MCMC sampler \texttt{emcee} \citep{emcee}. The MCMC was initialized with 120 walkers and evolved for 500,000 steps. We discarded the first 20\% of each chain as burn-in and thinned the remaining samples by a factor of 200. Convergence was assessed using the Gelman–Rubin statistic, yielding $\hat{R}<1.08$ for all parameters, and the total chain length exceeds at least 25 times the integrated autocorrelation time for every parameter \citep{GelmanRubin}.

\begin{figure}
    \centering
    \includegraphics[width=1.\linewidth]{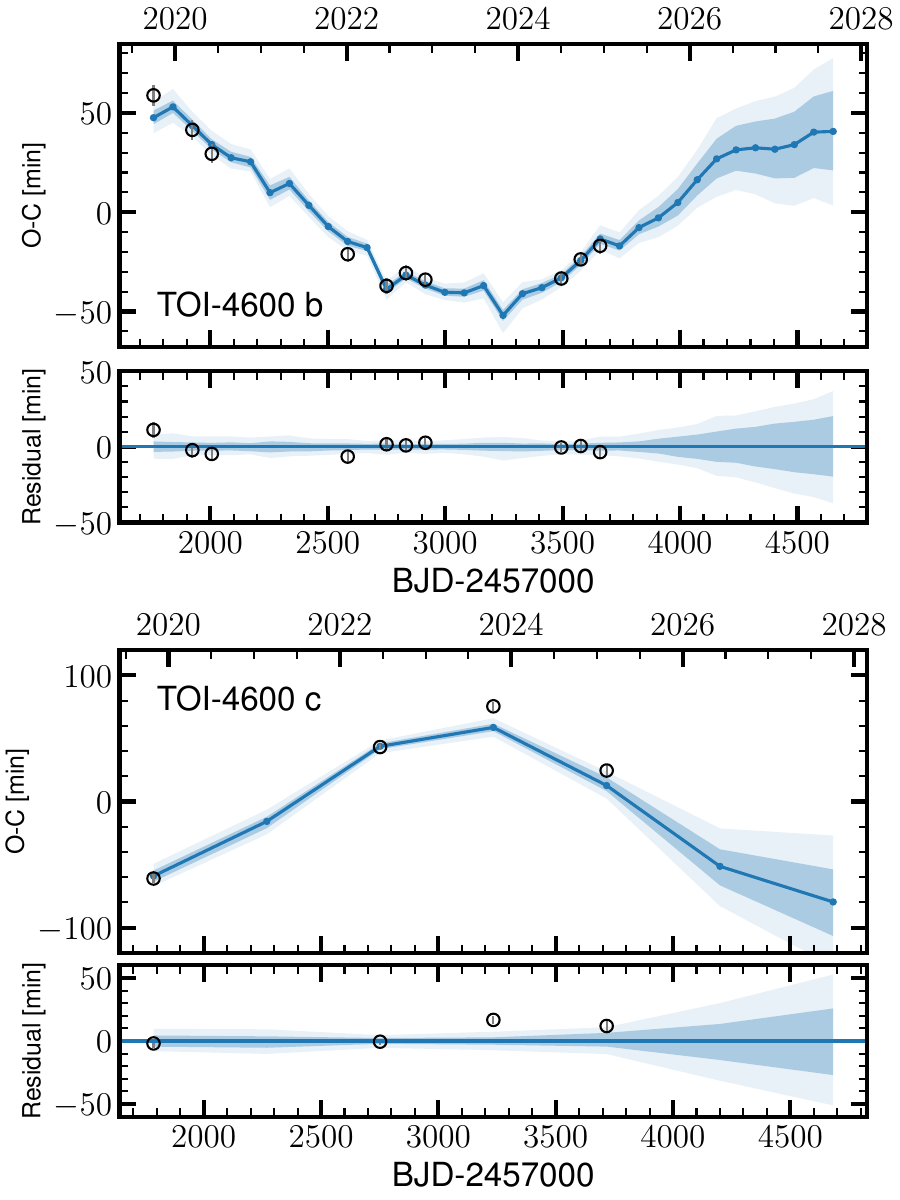}
    \caption{Transit timing variations of TOI-4600 b and c from the joint photodynamics and RV modeling. Open black circles are observed data. Blue lines are the median value of TTV predictions, with dark and light blue regions showing the 1 and 2$\sigma$ credible region. }
    \label{fig:ttv_prediction}
\end{figure}

The fitted and derived parameters of the photodynamical model are reported in Tables \ref{tab:fitted_parameter} and \ref{tab:derived_physical_parameter}, respectively. The TTV signals predicted from posterior samples are shown in Figure \ref{fig:ttv_prediction}.

\begin{deluxetable*}{lll}
\tablewidth{1.0\textwidth}
\tablecaption{Physical parameters for TOI-4600 b and c. 
Reported values are median with one $\sigma$ uncertainties; upper limits are quoted at the 95\% credible level. 
Osculating parameters are referenced to BJD = 2458745.0. \label{tab:derived_physical_parameter}}
\tablehead{\colhead{Parameter} & \colhead{TOI-4600 b} & \colhead{TOI-4600 c}  } 
\startdata
Orbital Period (d)                  & 
$82.712^{+0.0024}_{-0.0026}$      & 
$482.761^{+0.0065}_{-0.0064}$       \\
Planetary Mass (M$_\oplus$)         & $74.7^{+4.7}_{-4.4}$                  & $212.53^{+13.26}_{-13.03}$           \\ 
Planetary Radius (R$_\oplus$)       & $7.02^{+0.26}_{-0.21}$                & $9.78^{+0.38}_{-0.32}$               \\ 
Bulk Density (g/cm$^3$)             & $1.18^{+0.13}_{-0.13}$                & $1.25^{+0.15}_{-0.16}$               \\ 
Semi-major Axis (AU)                & $0.3485^{+0.0065}_{-0.0069}$          & $1.13^{+0.021}_{-0.022}$             \\ 
Eccentricity                        & $0.153^{+0.02}_{-0.018}$              & $0.219^{+0.015}_{-0.018}$            \\ 
Argument of Periastron (deg)        & $191.3^{+13.6}_{-12.9}$               & $102.6^{+6.8}_{-6.9}$               \\
Inclination (deg)                   & $89.81^{+0.099}_{-0.077}$             & $89.933^{+0.043}_{-0.031}$           \\ 
Scaled Semi-major Axis ($a/R_*$)    & $86.3^{+2.4}_{-2.7}$                  & $279.7^{+7.8}_{-8.9}$                \\ 
Mutual Inclination (deg)			&	-									& $7.3^{+7.7}_{-5.2}$ ($<20.1$)		   \\
\enddata
\end{deluxetable*}

\subsection{Adopted results}
The results from the transit fit, RV fit, and photodynamic fit are in agreement. We adopt the photodynamic fit as our preferred solution, with the key planetary parameters listed in Table~\ref{tab:derived_physical_parameter}.

\section{Discussion\label{section5}}

\subsection{TOI-4600 b,c among transiting giant planets}

\citet{Mireles2023} highlighted the distinctive position of TOI-4600 b and c in the period--equilibrium-temperature parameter space of known transiting planets, and identified TOI-4600 c as one of the longest-period giant planets known in a multi-transiting system. Our present work provides improved constraints on the planetary masses and orbital eccentricities. These refined parameters allow us to extend the contextual comparison in two important directions: the Transmission Spectroscopy Metric \citep[TSM;][]{Kempton2018_TSM} and orbital eccentricity.

\begin{figure*}[htbp]
    \centering
    \begin{subfigure}[b]{0.48\textwidth}
        \centering
        \includegraphics[width=\textwidth]{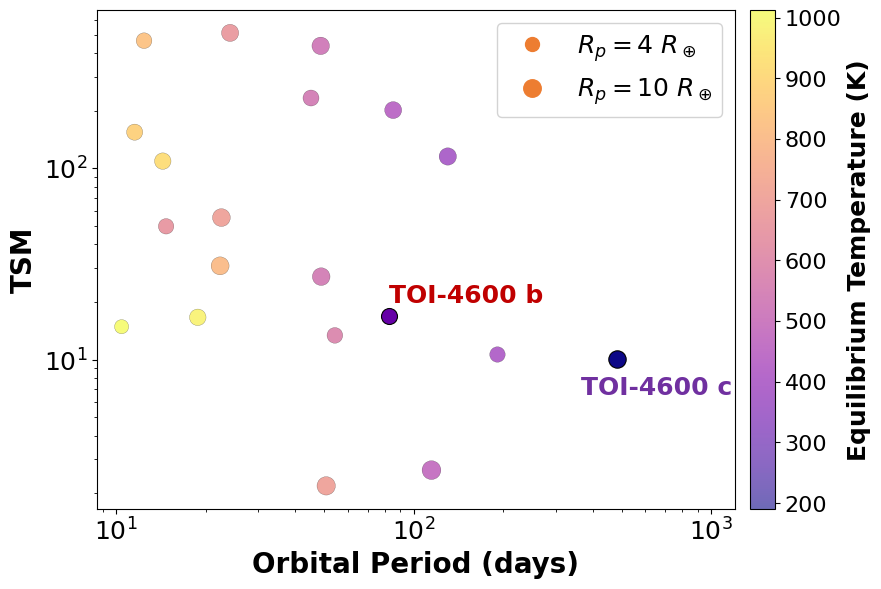}
        \label{fig:tsm_period}
    \end{subfigure}
    \hfill
    \begin{subfigure}[b]{0.48\textwidth}
        \centering
        \includegraphics[width=\textwidth]{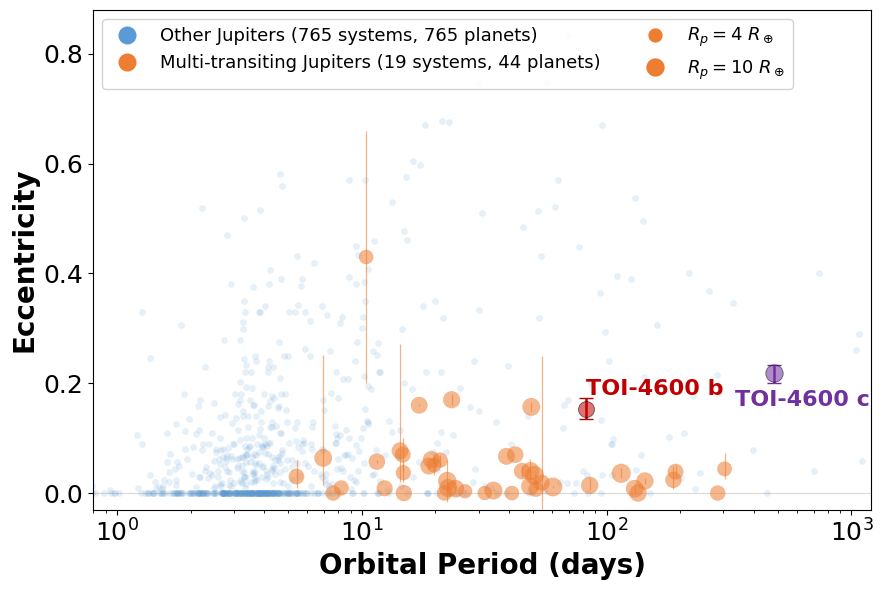}
        \label{fig:ecc_period}
    \end{subfigure}
    \caption{
    \textit{Left:} Transmission Spectroscopy Metric (TSM) versus orbital period for multi-transiting Jupiters (multiple transiting giant planets in one system). Giant planets are defined as $R_p \geq 4\,R_\oplus$. Circle size scales with planetary radius, and color indicates the equilibrium temperature evaluated at mid-transit.
    \textit{Right:} Orbital eccentricity versus orbital period for transiting giant planets. Multi-transiting Jupiters are shown as orange circles, and other giant planets are shown as light blue circles. Symbol size scales with planetary radius.
    }
    \label{fig:toi4600_context}
\end{figure*}

Figure~\ref{fig:toi4600_context} shows the TSM versus the orbital period of multi-transiting Jupiters (multiple transiting gas giants in one system). For consistency with the transit geometry, we computed the TSM using the equilibrium temperature evaluated at the star--planet separation at mid-transit, rather than the orbit-averaged equilibrium temperature. We obtained TSM values of 16.87 for TOI-4600\,b and 10.09 for TOI-4600\,c. While neither planet ranks among the highest-TSM targets in the sample, TOI-4600\,c is notable as a long-period, temperate transiting giant for which atmospheric characterization may be feasible, offering a rare opportunity to probe giant-planet atmospheres in a relatively cool and weakly irradiated regime.

Figure~\ref{fig:toi4600_context} also shows the eccentricity--period diagram of transiting giant planets ($R_p \geq 4\,R_\oplus$), with multi-transiting Jupiters highlighted. Both TOI-4600\,b and c exhibit moderate eccentricities of $e_b = 0.153^{+0.020}_{-0.018}$ and $e_c = 0.219^{+0.015}_{-0.018}$, respectively. Together with the improved mass measurements presented in this work, these eccentricity constraints make TOI-4600 one of the few long-period transiting giant-planet systems in the warm-Jupiter regime for which both planetary masses and eccentricities are precisely measured. Given their relatively large orbital separations, tidal circularization is expected to be inefficient, and the measured non-zero eccentricities may therefore preserve dynamical signatures of the system's formation and long-term evolution.

\subsection{Large TTVs in Long-period Multi-planet Systems}

At first glance, the large TTVs in TOI-4600, with semi-amplitudes of order $1$\,hr, may appear surprising because the two planets are not near a low-order mean-motion resonance. However, TOI-4600\,b and c lie close to the high-order \emph{6:1} MMR. Following the usual convention for measuring the distance from an exact commensurability \citep[e.g.,][]{Lithwick2012}, we define
\begin{equation}
    \Delta \equiv \frac{P_c/P_b}{6} - 1 .
\end{equation}
For TOI-4600, this gives $\Delta \simeq -2.7\%$, corresponding to a resonant super-period of
\begin{equation}
    P_{\rm sup} = \left| \frac{6}{P_c} - \frac{1}{P_b} \right|^{-1} \simeq 3000~{\rm days}.
\end{equation}
This near-6:1 configuration naturally produces an approximately sinusoidal and anti-correlated TTV pattern \citep{Deck_2016}, consistent with the behavior seen in Figure~\ref{fig:ttv_prediction}. In our photodynamical $N$-body solution, the relevant 6:1 resonant arguments circulate, rather than librate, on approximately the same super-period timescale.

Overall, the simulated transit timing variations in Figure~\ref{fig:ttv_prediction} reproduce the measured TTV signals of both TOI-4600\,b and TOI-4600\,c and capture the dominant periodicity. The last two TTV measurements of TOI-4600\,c show modest offsets relative to the median two-planet dynamical model. The origin of these offsets is currently unclear, but they could be related to underestimated timing uncertainties or residual systematics in the ground-based photometry.

Previous studies have shown that large TTVs can arise naturally when an inner transiting planet is perturbed by a massive outer companion on an eccentric and sufficiently separated orbit \citep{Holman_2005}. Although more complete analytic descriptions exist for TTVs near arbitrary-order mean-motion resonances \citep[e.g., Section~5 of][]{Deck_2016}, our goal here is not to provide a precise analytic model of the TOI-4600 TTV signal. Instead, we use the following expression as a simple order-of-magnitude estimate that can be readily applied by observers to identify similar systems likely to exhibit detectable TTVs. For an inner transiting planet perturbed by an exterior companion, the full TTV amplitude can be approximated as \citep[Eq.~1 of][]{Holman_2005}
\begin{equation}
    \Delta t \sim \frac{45\pi}{16}
    \left(\frac{M_2}{M_\star}\right)
    P_1 \alpha_e^3
    \left(1-\sqrt{2}\,\alpha_e^{3/2}\right)^{-2},
    \label{eq:hm2005_ttv}
\end{equation}
where
\begin{equation}
    \alpha_e = \frac{a_1}{a_2(1-e_2)} .
\end{equation}
Here, $P_1$ and $a_1$ are the orbital period and semi-major axis of the inner transiting planet, while $a_2$, $e_2$, and $M_2$ are the semi-major axis, eccentricity, and mass of the outer perturber. This expression is not intended to replace direct $N$-body modeling, but it provides useful physical intuition: the TTV amplitude increases with the perturber mass and depends strongly on the perturber's periastron distance through $\alpha_e$. For the measured parameters of TOI-4600, Equation~\ref{eq:hm2005_ttv} gives $\Delta t \simeq 2.0$\,hr for TOI-4600\,b, comparable to the observed and modeled signal.

For TOI-4600\,c, no similarly compact estimate applies in the same form because the perturbation is produced by the inner planet. Nevertheless, our $N$-body simulations show that the TTV amplitude of TOI-4600\,c scales approximately with the mass of TOI-4600\,b and with the orbital period of TOI-4600\,c. The large TTV amplitude of TOI-4600\,c is therefore physically expected in this system: even modest fractional perturbations to the orbital phase can correspond to large timing offsets for a long-period planet.

This interpretation is further supported by both previous studies and population statistics. In particular, \citet{Wang_2015} noted that long-period transiting planets frequently exhibit transit timing variations, suggesting that such systems commonly host additional dynamically interacting companions. We used the NASA Exoplanet Archive \citep{Christiansen2025}, accessed on 2026 April 4, to compile transiting planets in multi-planet systems and examined the occurrence of reported TTV signals. In the full sample, 388 of 1919 planets (\(\sim 20.2\%\)) show reported TTVs. Restricting the sample to planets with orbital periods \(>10\) days increases the fraction to 277 of 985 planets (\(\sim 28.1\%\)). Applying a further cut to planets with masses \(>20\,M_\oplus\) yields 96 planets, of which 61 (\(\sim 63.5\%\)) show reported TTV signals. Although these fractions are affected by observational selection effects and by heterogeneous reporting in the literature, they suggest that detectable TTVs are common among known long-period massive planets in multi-planet systems.

\subsection{Stellar rotation period and \texorpdfstring{$v_* \sin i$}{v sin i}}
\label{sec:vsini}

The low mutual inclination of TOI-4600 b and c suggests cold dynamical histories that may preserve their primordial obliquities. Future RM observations offer a unique opportunity to probe the primordial alignment of exoplanets. To assess the feasibility of such observations, we first sought to estimate the projected rotational velocity, $v_* \sin i$, following the methodology of \citep{Brady2024}. Using the slowly rotating K-star HD 32147 as a calibrator, we compared its MAROON-X spectrum to that of TOI-4600 via the cross-correlation function (CCF) method described in \cite{Gray2005}. We detected no rotational broadening signal for TOI-4600, indicating that its $v_* \sin i$ falls below the resolution limit of MAROON-X (approximately $< 2$ km/s). Assuming the stellar spin axis is aligned with the orbits of the transiting planets (i.e., $i_*=0^\circ$), this implies a stellar rotation period greater than 20 days.

To establish a lower limit for $v_* \sin i$, we attempted to predict the stellar rotation period using gyrochronology \citep{Barnes2003}. Because gyrochronology relations remain poorly calibrated for stars older than roughly 4 Gyr, we derived estimates based on several well-studied calibrators. K dwarfs in the open cluster M67 \citep[$\sim 4.2$ Gyr;][]{Barnes2016,Dungee2022,Long2023} exhibit rotation periods of around 30 days. The nearby K-dwarf benchmarks $\alpha$ Cen B \citep[$\sim 7.5$ Gyr;][]{Thevenin2026} and $\tau$ Ceti \citep[$\sim 9$ Gyr;][]{Tang2011} have rotation periods of roughly 37 days \citep{Dumusque2014} and 46 days \citep{Korolik2023}, respectively. Furthermore, HD 219134, a 10-Gyr K-star benchmark constrained by asteroseismology \citep{Li2025_HD219134}, has a rotation period of $\sim 41$ days. Given TOI-4600's estimated age of $3.9 \pm 2.2$ Gyr (isochrone fitting), we suggest a conservative upper limit of 50 days for the stellar rotation period, which corresponds to $v_* \sin i = 0.8$ km/s when assuming $i_*=0^\circ$.

Adopting $v_* \sin i = 0.8$ km/s, the predicted RM amplitudes \citep{Albrecht2022} are 2.9 m/s and 5.8 m/s for TOI-4600 b and c, respectively, both of which are measurable by current high-precision RV instruments. Even if the true signal is smaller than anticipated due to a non-zero $i_*$, combining the resulting upper limit on $v_* \sin i$ with the rotation period constraints will still yield valuable insights into the primordial misalignment of this system (Section~\ref{sec:intro}).

\section{Summary and Future Work\label{section6}}
In this work, we refine the planetary parameters of TOI-4600 b and c using new TESS observations, ground-based transit follow-up, and high-precision RVs from MAROON-X. For TOI-4600\,b, we derive an orbital period of $82.7$ days, a mass of $74.7^{+4.7}_{-4.4}\,M_{\oplus}$, a radius of $7.02^{+0.26}_{-0.21}\,R_{\oplus}$, and an eccentricity of $0.153^{+0.020}_{-0.018}$. For TOI-4600\,c, we derive an orbital period of $482.8$ days, a mass of $212.53^{+13.26}_{-13.03}\,M_{\oplus}$, a radius of $9.78^{+0.38}_{-0.32}\,R_{\oplus}$, and an eccentricity of $0.219^{+0.015}_{-0.018}$. We detect significant TTVs in both planets with semi-amplitudes of approximately $1$ hr, and we provide predicted TTV ephemerides for future events in Table \ref{tab:epoch_params_b_and_c_future}. 

We estimate Transit Spectroscopy Metric (TSM) values of $16.87$ and $10.09$ for TOI-4600 b and c, respectively. These metrics highlight that both planets are rare and favorable targets for future JWST atmospheric studies of temperate giant planets. The small mutual inclination suggests that strong planet--planet scattering is unlikely to have played a major role in shaping the current architecture. The stellar obliquity of TOI-4600 therefore likely reflects the primordial misalignment of the protoplanetary disk, and future Rossiter--McLaughlin observations will be crucial for directly constraining this scenario.

\begin{acknowledgments}
TH, SXW, ZL acknowledge support from NSFC grant 12273016. JZ and SXW acknowledge support from MOST grant 2025YFE0102100. We thank Zhecheng Hu for his valuable suggestions on dynamic and TTV analysis.

We thank Arthur Sweeney and Nicholas Dowling for their help in making the SOFAR and Wendelstein observations, respectively, available through ExoFOP.

This work is partially based on observations obtained at the international Gemini Observatory, a program of NSF NOIRLab, which is managed by the Association of Universities for Research in Astronomy (AURA) under a cooperative agreement with the U.S. National Science Foundation on behalf of the Gemini Observatory partnership: the U.S. National Science Foundation (United States), National Research Council (Canada), Agencia Nacional de Investigación y Desarrollo (Chile), Ministerio de Ciencia, Tecnología e Innovación (Argentina), Ministério da Ciência, Tecnologia, Inovações e Comunicações (Brazil), and Korea Astronomy and Space Science Institute (Republic of Korea). This work was enabled by observations made from the Gemini North telescope, located within the Maunakea Science Reserve and adjacent to the summit of Maunakea. We are grateful for the privilege of observing the Universe from a place that is unique in both its astronomical quality and its cultural significance.

The MAROON-X team acknowledges funding from the David and Lucile Packard Foundation, the Heising-Simons Foundation, the Gordon and Betty Moore Foundation, the Gemini Observatory, the NSF (award number 2108465), and NASA (grant number 80NSSC22K0117).

The \textit{Tierras} Observatory is supported by the National Science Foundation under Award No.\ AST-2308043. J.G.-M.\ acknowledges support from the Heising-Simons Foundation, and from the Pappalardo family through the MIT Pappalardo Fellowship in Physics.

This work made use of juliet, a community-developed PYTHON package for the joint modeling of exoplanet photometry and radial velocities \citep{Foreman-Mackey_2017}. This research also made use of LIGHTKURVE, a PYTHON package for Kepler and TESS data analysis \citep{Lightkurve}, and REBOUND, an open-source multi-purpose N-body code \citep{Rein_2012}.

N.Chazov acknowledges support from the youth scientific laboratory project, topic FEUZ-2025-0003.

The work of HPO has been carried out within the framework of the NCCR PlanetS supported by the Swiss National Science Foundation under grants 51NF40\_182901 and 51NF40\_205606.

J.Z.L. was supported by the Tianshan Talent Training Program through the grant 2023TSYCCX0101, and the National Natural Science Foundation of China (grant Nos. 12433007 and 12541303), the Urumqi Nanshan Astronomy and Deep Space Exploration Observation and Research Station of Xinjiang through the grant XJYWZ2303. Nanshan 1m Telescope is partly supported by the Operation, Maintenance and Upgrading Fund for Astronomical Telescopes and Facility Instruments,budgeted from the Ministry of Finance of China (MOF) and administrated by the Chinese Academy of Sciences (CAS).

\end{acknowledgments}

\bibliography{sample701}{}
\bibliographystyle{aasjournalv7}

\clearpage
\onecolumngrid
\appendix

\section{Predicted Transit Times}
\label{app:predicted_transit_times}

\begin{table}[!htbp]
\centering
\caption{
Predicted future mid-transit times for TOI-4600 b and TOI-4600 c.
Times are given in BJD. The timing uncertainty, $\sigma_{t_0}$, is taken as the average of the lower and upper 1$\sigma$ bounds.
}
\label{tab:epoch_params_b_and_c_future}
\begin{tabular}{lccc}
\hline
Planet & Epoch & $t_0$ (BJD) & $\sigma_{t_0}$ (d) \\
\hline
b & 1  & 2461073.2452 & 0.0057 \\
b & 2  & 2461155.9439 & 0.0073 \\
b & 3  & 2461238.6396 & 0.0075 \\
b & 4  & 2461321.3322 & 0.0087 \\
b & 5  & 2461404.0238 & 0.0096 \\
b & 6  & 2461486.7170 & 0.0105 \\
b & 7  & 2461569.4130 & 0.0120 \\
b & 8  & 2461652.1050 & 0.0140 \\
b & 9  & 2461734.8010 & 0.0135 \\
b & 10 & 2461817.4870 & 0.0140 \\
b & 11 & 2461900.1730 & 0.0145 \\
b & 12 & 2461982.8610 & 0.0145 \\
\hline
c & 1  & 2461199.8870 & 0.0105 \\
c & 2  & 2461682.6520 & 0.0180 \\
\hline
\end{tabular}
\end{table}

\FloatBarrier
\clearpage

\section{MAROON-X RV Data}
\label{app:maroonx_rv_data}

\begin{table}[!htbp]
\centering
\caption{
MAROON-X radial velocities from the red and blue arms.
}
\label{tab:maroonx_rv_combined}
\begin{tabular}{ccccc}
\hline
\hline
Time (BJD) &
RV$_{\rm red}$ &
$\sigma_{\rm red}$ &
RV$_{\rm blue}$ &
$\sigma_{\rm blue}$ \\
&
(m\,s$^{-1}$) &
(m\,s$^{-1}$) &
(m\,s$^{-1}$) &
(m\,s$^{-1}$) \\
\hline
2460369.0952 & -10.89 &  2.39 & -13.02 & 1.53 \\
2460386.0272 & -12.88 &  4.34 &  -2.27 & 2.40 \\
2460435.0182 & -14.90 &  2.21 & -13.67 & 1.19 \\
2460441.1127 & -12.35 &  3.72 &  -8.97 & 2.12 \\
2460451.8736 &  12.11 &  2.92 &   8.35 & 1.62 \\
2460459.0212 &   8.99 &  3.07 &   8.55 & 1.68 \\
2460459.9815 &  16.89 &  3.77 &  13.71 & 2.04 \\
2460483.8702 &  23.62 &  3.38 &  13.70 & 1.86 \\
2460484.8354 &  15.96 &  2.18 &  11.92 & 1.20 \\
2460485.9940 &   9.15 &  3.59 &  17.07 & 1.99 \\
2460490.8991 &   6.89 &  2.30 &   6.98 & 1.28 \\
2460497.9307 &  10.59 &  2.73 &   2.91 & 1.52 \\
2460503.9229 &  -7.11 & 10.10 &   0.94 & 5.14 \\
2460504.8872 &   4.60 &  4.34 &  -0.98 & 2.37 \\
2460505.8231 &   4.59 &  5.42 &  -0.97 & 2.93 \\
2460507.8652 &  -7.23 &  3.62 &  -0.89 & 2.03 \\
2460516.8215 &  10.91 &  6.30 &   1.62 & 3.17 \\
2460517.8740 &   2.61 &  2.97 &  -0.95 & 1.62 \\
2460521.8356 &   8.78 &  2.17 &   4.27 & 1.17 \\
2460407.9749 & -15.79 &  3.01 &  -4.34 & 1.65 \\
2460411.0431 &  -7.65 &  3.52 &  -7.19 & 1.94 \\
2460424.1346 & -21.51 &  2.17 & -16.46 & 1.17 \\
\hline
\end{tabular}
\end{table}

\FloatBarrier
\clearpage

\section{Transit Fit Result}
\label{app:transit_fit_result}

\startlongtable
\begin{deluxetable}{llll}
\tabletypesize{\scriptsize}
\tablewidth{1.0\textwidth}
\tablecaption{
Fitted parameters for the transit modeling of TOI-4600.
Reported values are medians with 1$\sigma$ uncertainties.
$\mathcal{N}(\mu,\sigma)$ denotes a normal prior with mean $\mu$ and standard deviation $\sigma$, and $\mathcal{U}(a,b)$ denotes a uniform prior between $a$ and $b$.
\label{tab:fitted-params-transit}
}
\tablehead{
\colhead{Parameter} &
\colhead{Description} &
\colhead{Prior} &
\colhead{Posterior}
}
\startdata
\multicolumn{4}{c}{\textbf{Planet b}}\\
\hline
\multicolumn{4}{l}{\textit{Stellar parameter}}\\
$\rho_\star$ [kg\,m$^{-3}$] & Stellar density & $\mathcal{N}(2270.0,320.0)$ & $2236^{+268}_{-245}$ \\
\hline
\multicolumn{4}{l}{\textit{Planet b --- fitted parameters}}\\
$r_1$       & Transit-shape parameter & $\mathcal{U}(0.0,1.0)$ & $0.56^{+0.13}_{-0.14}$ \\
$r_2$       & Radius-ratio parameter & $\mathcal{U}(0.0,0.2)$ & $0.0750^{+0.0025}_{-0.0017}$ \\
$\sqrt{e}\sin\omega$  & Eccentricity component & $\mathcal{U}(-0.6,0.6)$ & $-0.30^{+0.13}_{-0.12}$ \\
$\sqrt{e}\cos\omega$  & Eccentricity component & $\mathcal{U}(-0.6,0.6)$ & $-0.028^{+0.396}_{-0.374}$ \\
\hline
\multicolumn{4}{l}{\textit{Planet b --- derived parameters}}\\
$P$ [d] & Orbital period & --- & $82.68952^{+0.00012}_{-0.00011}$ \\
$t_0$ [BJD$-$2457000] & Transit epoch & --- & $1757.8844^{+0.0018}_{-0.0019}$ \\
$p = R_{\rm p}/R_\star$  & Radius ratio & --- & $0.0750^{+0.0025}_{-0.0017}$ \\
$b$  & Impact parameter & --- & $0.34^{+0.19}_{-0.21}$ \\
$i$ [$^\circ$] & Inclination & --- & $89.81^{+0.11}_{-0.09}$ \\
$e$  & Eccentricity & --- & $0.19^{+0.15}_{-0.10}$ \\
$\omega$ [$^\circ$] & Argument of periastron & --- & $-91.8^{+58.5}_{-52.7}$ \\
$a/R_\star$  & Scaled semi-major axis & --- & $93.1^{+3.6}_{-3.5}$ \\
\hline
\multicolumn{4}{l}{\textit{TESS photometric parameters}}\\
$q_{0,\rm TESS}$  & Limb-darkening coefficient & $\mathcal{U}(0.0,1.0)$ & $0.58^{+0.26}_{-0.27}$ \\
$q_{1,\rm TESS}$  & Limb-darkening coefficient & $\mathcal{U}(0.0,1.0)$ & $0.23^{+0.23}_{-0.12}$ \\
$m_{{\rm dilution},\rm TESS}$  & Dilution factor & fixed ($1.0$) & $\equiv 1.0$ \\
$m_{{\rm flux},\rm TESS}$  & Flux offset & $\mathcal{N}(0.0,0.1)$ & $-0.47^{+1.3}_{-1.4}\times10^{-5}$ \\
$\sigma_{w,\rm TESS}$  & Photometric jitter & $\log\mathcal{U}(0.1,1000.0)$ & $1.8^{+12.5}_{-1.6}$ \\
\hline
\multicolumn{4}{c}{\textbf{Planet c}}\\
\hline
\multicolumn{4}{l}{\textit{Stellar parameter}}\\
$\rho_\star$ [kg\,m$^{-3}$] & Stellar density & $\mathcal{N}(2270.0,320.0)$ & $2459^{+283}_{-305}$ \\
\hline
\multicolumn{4}{l}{\textit{Planet c --- fitted parameters}}\\
$P$ [d] & Orbital period & $\mathcal{U}(481.82,483.82)$ & $482.8196^{+0.0019}_{-0.0017}$ \\
$t_0$ [BJD$-$2457000] & Transit epoch & $\mathcal{U}(2750.6008,2752.6008)$ & $2751.6010^{+0.0020}_{-0.0020}$ \\
$r_1$  & Transit-shape parameter & $\mathcal{U}(0.0,1.0)$ & $0.634^{+0.080}_{-0.135}$ \\
$r_2$  & Radius-ratio parameter & $\mathcal{U}(0.0,0.2)$ & $0.1061^{+0.0029}_{-0.0029}$ \\
$\sqrt{e}\sin\omega$  & Eccentricity component & $\mathcal{U}(-0.6,0.6)$ & $0.14^{+0.18}_{-0.22}$ \\
$\sqrt{e}\cos\omega$  & Eccentricity component & $\mathcal{U}(-0.6,0.6)$ & $-0.08^{+0.45}_{-0.34}$ \\
\hline
\multicolumn{4}{l}{\textit{Planet c --- derived parameters}}\\
$p = R_{\rm p}/R_\star$  & Radius ratio & --- & $0.1061^{+0.0029}_{-0.0029}$ \\
$b$  & Impact parameter & --- & $0.45^{+0.12}_{-0.20}$ \\
$i$ [$^\circ$] & Inclination & --- & $89.911^{+0.035}_{-0.018}$ \\
$e$  & Eccentricity & --- & $0.16^{+0.11}_{-0.10}$ \\
$\omega$ [$^\circ$] & Argument of periastron & --- & $60^{+93}_{-140}$ \\
$a/R_\star$  & Scaled semi-major axis & --- & $312^{+12}_{-13}$ \\
\hline
\multicolumn{4}{l}{\textit{TESS photometric parameters}}\\
$q_{0,\rm TESS}$  & Limb-darkening coefficient & $\mathcal{U}(0.0,1.0)$ & $0.35^{+0.34}_{-0.17}$ \\
$q_{1,\rm TESS}$  & Limb-darkening coefficient & $\mathcal{U}(0.0,1.0)$ & $0.34^{+0.34}_{-0.21}$ \\
$m_{{\rm dilution},\rm TESS}$  & Dilution factor & fixed ($1.0$) & $\equiv 1.0$ \\
$m_{{\rm flux},\rm TESS}$  & Flux offset & $\mathcal{N}(0.0,0.1)$ & $-3.9\times10^{-5}{}^{+3.3\times10^{-5}}_{-3.4\times10^{-5}}$ \\
$\sigma_{w,\rm TESS}$  & Photometric jitter & $\log\mathcal{U}(0.1,1000.0)$ & $2.3^{+23.7}_{-2.1}$ \\
\enddata
\end{deluxetable}

\FloatBarrier
\clearpage

\section{RV Fit Result}
\label{app:rv_fit_result}

\startlongtable
\begin{deluxetable}{llll}
\tabletypesize{\scriptsize}
\tablewidth{1.0\textwidth}
\tablecaption{
Fitted parameters for the RV modeling of TOI-4600.
Reported values are medians with 1$\sigma$ uncertainties.
$\mathcal{N}(\mu,\sigma)$ denotes a normal prior with mean $\mu$ and standard deviation $\sigma$, and $\mathcal{U}(a,b)$ denotes a uniform prior between $a$ and $b$.
The \texttt{blue} and \texttt{red} RV datasets correspond to the MAROON-X blue and red arms, respectively.
The $M\sin i$ values were computed assuming a stellar mass of $0.88\,M_\odot$ from our stellar characterization.
\label{tab:rv-fitted-params}
}
\tablehead{
\colhead{Parameter} &
\colhead{Description} &
\colhead{Prior} &
\colhead{Posterior}
}
\startdata
\multicolumn{4}{c}{\textbf{Instrument and systemic parameters}}\\
\hline
$\gamma_{\rm red}$ [m\,s$^{-1}$] & Systemic RV offset for the red arm & $\mathcal{U}(-50,50)$ & $0.8^{+1.8}_{-2.3}$ \\
$\gamma_{\rm blue}$ [m\,s$^{-1}$] & Systemic RV offset for the blue arm & $\mathcal{U}(-50,50)$ & $0.07^{+1.63}_{-2.21}$ \\
$\sigma_{w,{\rm red}}$ [m\,s$^{-1}$] & Additional white-noise jitter for the red arm & $\log\mathcal{U}(0.001,100.0)$ & $2.6^{+1.3}_{-2.4}$ \\
$\sigma_{w,{\rm blue}}$ [m\,s$^{-1}$] & Additional white-noise jitter for the blue arm & $\log\mathcal{U}(0.001,100.0)$ & $0.20^{+0.86}_{-0.19}$ \\
\hline
\multicolumn{4}{c}{\textbf{Planet b --- RV Keplerian model}}\\
\hline
$P$ [d] & Orbital period & fixed ($82.69$) & $\equiv 82.69$ \\
$t_0$ [BJD$_{\rm TDB}$] & Time of inferior conjunction & fixed ($2460494.37$) & $\equiv 2460494.37$ \\
$K$ [m\,s$^{-1}$] & RV semi-amplitude & $\mathcal{U}(0.0,100.0)$ & $12.10^{+0.57}_{-0.58}$ \\
$\sqrt{e}\cos\omega$  & Eccentricity component & $\mathcal{N}(0.03,0.38)$ & $-0.362^{+0.031}_{-0.023}$ \\
$\sqrt{e}\sin\omega$  & Eccentricity component & $\mathcal{N}(-0.3,0.13)$ & $-0.153^{+0.070}_{-0.070}$ \\
\hline
\multicolumn{4}{l}{\textit{Planet b --- derived parameters}}\\
$M\sin i$ [$M_{\oplus}$] & Minimum mass inferred from RV modeling & --- & $74.8^{+3.5}_{-3.6}$ \\
$e$  & Orbital eccentricity & --- & $0.158^{+0.025}_{-0.022}$ \\
$\omega$ [$^\circ$] & Argument of periastron & --- & $-157^{+10}_{-10}$ \\
\hline
\multicolumn{4}{c}{\textbf{Planet c --- RV Keplerian model}}\\
\hline
$P$ [d] & Orbital period & fixed ($482.82$) & $\equiv 482.82$ \\
$t_0$ [BJD$_{\rm TDB}$] & Time of inferior conjunction & fixed ($2459751.60$) & $\equiv 2459751.60$ \\
$K$ [m\,s$^{-1}$] & RV semi-amplitude & $\mathcal{U}(0.0,100.0)$ & $20.3^{+7.0}_{-3.5}$ \\
$\sqrt{e}\cos\omega$  & Eccentricity component & $\mathcal{N}(-0.08,0.39)$ & $-0.342^{+0.091}_{-0.081}$ \\
$\sqrt{e}\sin\omega$  & Eccentricity component & $\mathcal{N}(0.14,0.2)$ & $0.32^{+0.16}_{-0.19}$ \\
\hline
\multicolumn{4}{l}{\textit{Planet c --- derived parameters}}\\
$M\sin i$ [$M_{\oplus}$] & Minimum mass inferred from RV modeling & --- & $224^{+64}_{-36}$ \\
$e$  & Orbital eccentricity & --- & $0.22^{+0.14}_{-0.08}$ \\
$\omega$ [$^\circ$] & Argument of periastron & --- & $134^{+19}_{-19}$ \\
\enddata
\end{deluxetable}

\FloatBarrier
\clearpage

\section{Photodynamic Fit Result}
\label{app:photodynamic_fit_result}

\startlongtable
\begin{deluxetable}{llll}
\tabletypesize{\scriptsize}
\tablewidth{1.0\textwidth}
\tablecaption{
Fitted parameters for the joint photodynamical and RV modeling of TOI-4600.
Reported values are medians with 1$\sigma$ uncertainties.
Osculating parameters are defined in Jacobian coordinates and are referenced to BJD = 2458745.0.
$\mathcal{N}(\mu,\sigma)$ denotes a normal prior with mean $\mu$ and standard deviation $\sigma$, and $\mathcal{U}(a,b)$ denotes a uniform prior between $a$ and $b$.
\label{tab:fitted_parameter}
}
\tablehead{
\colhead{Parameter} &
\colhead{Prior} &
\colhead{Posterior}
}
\startdata
\multicolumn{3}{c}{\textbf{Stellar parameters}}\\
\hline
$R_*$ [$R_\odot$] & $\mathcal{N}(0.817,0.035)$ & $0.869^{+0.025}_{-0.023}$ \\
$M_*$ [$M_\odot$] & $\mathcal{N}(0.88,0.05)$ & $0.825^{+0.047}_{-0.048}$ \\
\hline
\multicolumn{3}{c}{\textbf{Planet b}}\\
\hline
$M_b/M_*$ & $\mathcal{U}(10^{-6},10^{-3})$ & $0.000272^{+1.4\times10^{-5}}_{-1.3\times10^{-5}}$ \\
$R_b/R_*$ & $\mathcal{N}(0.0769,0.005)$ & $0.07408^{+0.00119}_{-0.00089}$ \\
$P_b$ [d] & $\mathcal{U}(82.67,82.69)$ & $82.7123^{+0.0024}_{-0.0026}$ \\
$e_b\cos\omega_b$ & $\mathcal{U}(-0.5,0.5)$ & $-0.146^{+0.017}_{-0.017}$ \\
$e_b\sin\omega_b$ & $\mathcal{U}(-0.5,0.5)$ & $-0.029^{+0.033}_{-0.038}$ \\
$b_b$ & $\mathcal{U}(0,0.9)$ & $0.29^{+0.12}_{-0.16}$ \\
$\Omega_b$ & fixed & $\equiv 0$ \\
$T_{\rm conj,b}$ [BJD$-$2457000] & $\mathcal{U}(1757.8,1758.0)$ & $1757.893^{+0.0028}_{-0.0027}$ \\
\hline
\multicolumn{3}{c}{\textbf{Planet c}}\\
\hline
$M_c/M_*$ & $\mathcal{U}(10^{-6},10^{-3})$ & $0.000775^{+3.9\times10^{-5}}_{-3.8\times10^{-5}}$ \\
$R_c/R_*$ & $\mathcal{N}(0.1067,0.005)$ & $0.1032^{+0.0015}_{-0.0013}$ \\
$P_c$ [d] & $\mathcal{U}(482.62,483.02)$ & $482.7607^{+0.0065}_{-0.0064}$ \\
$e_c\cos\omega_c$ & $\mathcal{U}(-0.5,0.5)$ & $-0.048^{+0.027}_{-0.029}$ \\
$e_c\sin\omega_c$ & $\mathcal{U}(-0.5,0.5)$ & $0.212^{+0.011}_{-0.014}$ \\
$b_c$ & $\mathcal{U}(0,0.9)$ & $0.26^{+0.11}_{-0.16}$ \\
$\Omega_c$ & $\mathcal{U}(-\pi/5,\pi/5)$ & $0.03^{+0.17}_{-0.19}$ \\
$T_{\rm conj,c}$ [BJD$-$2457000] & $\mathcal{U}(1785.86,1786.06)$ & $1785.9653^{+0.0030}_{-0.0029}$ \\
\hline
\multicolumn{3}{c}{\textbf{RV parameters}}\\
\hline
$\gamma_{\rm blue}$ [m\,s$^{-1}$] & $\mathcal{U}(-20,20)$ & $3.5^{+1.0}_{-1.0}$ \\
$\gamma_{\rm red}$ [m\,s$^{-1}$] & $\mathcal{U}(-20,20)$ & $4.3^{+1.2}_{-1.2}$ \\
$\log \sigma_{\rm blue}$ & $\mathcal{U}(-4,4)$ & $-0.31^{+0.84}_{-2.36}$ \\
$\log \sigma_{\rm red}$ & $\mathcal{U}(-4,4)$ & $0.76^{+0.50}_{-2.34}$ \\
\hline
\multicolumn{3}{c}{\textbf{Photometric parameters}}\\
\hline
$q_{\rm 0,TESS}$ & $\mathcal{U}(0,1)$ & $0.79^{+0.14}_{-0.21}$ \\
$q_{\rm 1,TESS}$ & $\mathcal{U}(0,1)$ & $0.154^{+0.098}_{-0.072}$ \\
$\log \sigma_{\rm TESS,2min}$ & $\mathcal{U}(-9,-3)$ & $-8.59^{+0.40}_{-0.29}$ \\
$\log \sigma_{\rm TESS,30min}$ & $\mathcal{U}(-9,-3)$ & $-8.63^{+0.39}_{-0.26}$ \\
$q_{\rm 0,NEOSAT}$ & $\mathcal{U}(0,1)$ & $0.083^{+0.084}_{-0.052}$ \\
$q_{\rm 1,NEOSAT}$ & $\mathcal{U}(0,1)$ & $0.61^{+0.27}_{-0.35}$ \\
$\log \sigma_{\rm NEOSAT}$ & $\mathcal{U}(-9,-3)$ & $-6.54^{+0.38}_{-1.09}$ \\
$q_{\rm 0,nanshan}$ & $\mathcal{U}(0,1)$ & $0.34^{+0.32}_{-0.21}$ \\
$q_{\rm 1,nanshan}$ & $\mathcal{U}(0,1)$ & $0.30^{+0.35}_{-0.21}$ \\
$\log \sigma_{\rm nanshan}$ & $\mathcal{U}(-9,-3)$ & $-6.052^{+0.073}_{-0.073}$ \\
$q_{\rm 0,Kourovka,R}$ & $\mathcal{N}(0.50,0.2)$ & $0.65^{+0.14}_{-0.13}$ \\
$q_{\rm 1,Kourovka,R}$ & $\mathcal{N}(0.36,0.2)$ & $0.54^{+0.14}_{-0.14}$ \\
$\log \sigma_{\rm Kourovka,R}$ & $\mathcal{U}(-9,-3)$ & $-6.2^{+0.19}_{-0.24}$ \\
$q_{\rm 0,Kourovka,V}$ & $\mathcal{N}(0.59,0.2)$ & $0.73^{+0.14}_{-0.15}$ \\
$q_{\rm 1,Kourovka,V}$ & $\mathcal{N}(0.41,0.2)$ & $0.54^{+0.14}_{-0.14}$ \\
$\log \sigma_{\rm Kourovka,V}$ & $\mathcal{U}(-9,-3)$ & $-5.84^{+0.16}_{-0.19}$ \\
$q_{\rm 0,Sofar}$ & $\mathcal{N}(0.67,0.2)$ & $0.54^{+0.16}_{-0.15}$ \\
$q_{\rm 1,Sofar}$ & $\mathcal{N}(0.46,0.2)$ & $0.52^{+0.19}_{-0.19}$ \\
$\log \sigma_{\rm Sofar}$ & $\mathcal{U}(-9,-3)$ & $-5.7^{+0.11}_{-0.12}$ \\
$q_{\rm 0,StAndrews}$ & $\mathcal{U}(0,1)$ & $0.85^{+0.11}_{-0.17}$ \\
$q_{\rm 1,StAndrews}$ & $\mathcal{U}(0,1)$ & $0.914^{+0.064}_{-0.133}$ \\
$\log \sigma_{\rm StAndrews}$ & $\mathcal{U}(-9,-3)$ & $-4.993^{+0.064}_{-0.064}$ \\
$q_{\rm 0,Tierras}$ & $\mathcal{N}(0.40,0.2)$ & $0.9868^{+0.0098}_{-0.0214}$ \\
$q_{\rm 1,Tierras}$ & $\mathcal{N}(0.33,0.2)$ & $0.98^{+0.015}_{-0.035}$ \\
$\log \sigma_{\rm Tierras}$ & $\mathcal{U}(-9,-3)$ & $-6.53^{+0.18}_{-0.24}$ \\
$q_{\rm 0,WST}$ & $\mathcal{N}(0.25,0.2)$ & $0.088^{+0.137}_{-0.069}$ \\
$q_{\rm 1,WST}$ & $\mathcal{N}(0.22,0.2)$ & $0.23^{+0.18}_{-0.15}$ \\
$\log \sigma_{\rm WST}$ & $\mathcal{U}(-9,-3)$ & $-4.739^{+0.060}_{-0.060}$ \\
\enddata
\end{deluxetable}



\end{document}